\documentclass[twocolumn,tighten,twocolappendix]{aastex631}
\usepackage{apjfonts}
\usepackage{amssymb}
\usepackage{amsbsy}
\usepackage{amsmath}
\usepackage{amsfonts}
\usepackage{mathrsfs}
\usepackage{color}
\usepackage{bm}
\usepackage{ulem}

\def\ihep{Key Laboratory for Particle Astrophysics, Institute of High Energy Physics, Chinese Academy of Sciences, 19B Yuquan Road, Beijing 100049, China; \href{mailto:liyanrong@ihep.ac.cn}{liyanrong@mail.ihep.ac.cn}, \href{mailto:wangjm@mail.ihep.ac.cn}{wangjm@mail.ihep.ac.cn}}

\def\UCASphys{School of Physics, University of Chinese Academy of Sciences, Beijing 100049, China}

\def\UCASastro{School of Astronomy and Space Sciences, University of Chinese Academy of Sciences, Beijing 100049, China}

\def\NAOC{National Astronomical Observatories of China, Chinese Academy of Sciences, 20A Datun Road, Beijing 100020, China}

\def\DongGuan{Dongguan Neutron Science Center, 1 Zhongziyuan Road, Dongguan 523808, China}

\def\YNAO{Yunnan Observatories, Chinese Academy of Sciences, Kunming 650011, China}

\def\KIAA{The Kavli Institute for Astronomy and Astrophysics, Peking University, Beijing 100871, China}
\def\PKUDoA{Department of Astronomy, School of Physics, Peking University, Beijing 100871, China}

\def\SAfrica{Department of Physics, University of Johannesburg, P.O. Box 524, 2006 Auckland Park, Johannesburg, South Africa}

\def\CAHA{Centro Astronomico Hispano Alem\'an, Sierra de los filabres sn, 04550 gergal. Almer\'ia, Spain}
\def\IAA{Instituto de Astrof\'isica de Andaluc\'ia, Glorieta de la astronom\'ia sn,
18008 Granada, Spain}

\def\Langrange{Universit\'e Côte d'Azur, Observatoire de la C\^ote d'Azur, CNRS, Laboratoire Lagrange UMR 7293, B\^atiment H. Fizeau, F-06108 Nice Cedex 2, France}

\begin{document}
\title{\bf\large Spectroastrometry and Reverberation Mapping (SARM) of Active Galactic Nuclei. \\ \vglue 0.2cm
I. The H$\beta$ Broad-line Region Structure and Black Hole Mass of Five Quasars}

\author[0000-0001-5841-9179]{Yan-Rong Li}
\author{Chen Hu}
   \affiliation{\ihep}
\author{Zhu-Heng Yao}
   \affiliation{\ihep}\affiliation{\UCASphys}
\author{Yong-Jie Chen}
   \affiliation{\ihep}
   \affiliation{\DongGuan}
\author{Hua-Rui Bai}
   \author{Sen Yang}
   \affiliation{\ihep}
   \affiliation{\UCASphys}
\author[0000-0002-5830-3544]{Pu Du}
   \affiliation{\ihep}
\author{Feng-Na Fang}
\author{Yi-Xin Fu}
\author{Jun-Rong Liu}
\author{Yue-Chang Peng}
   \affiliation{\ihep}
   \affil{\UCASphys}
\author[0000-0003-4042-7191]{Yu-Yang Songsheng}
   \affiliation{\ihep}
   \affiliation{\DongGuan}
\author{Yi-Lin Wang}
   \affiliation{\ihep}
   \affil{\UCASphys}
\author[0000-0001-5981-6440]{Ming Xiao}
   \affiliation{\ihep}
\author{Shuo Zhai}
   \affiliation{\ihep}
   \affil{\UCASphys}
\author[0000-0003-2662-0526]{Hartmut Winkler}
   \affil{\SAfrica}
\author{Jin-Ming Bai}
   \affiliation{\YNAO}
\author[0000-0001-6947-5846]{Luis C. Ho}
   \affiliation{\KIAA}
   \affiliation{\PKUDoA}
\author[0000-0003-4759-6051]{Romain G. Petrov}
   \affiliation{\Langrange}
\author{Jes\'us Aceituno}
   \affiliation{\CAHA}
   \affiliation{\IAA}
\author[0000-0001-9449-9268]{Jian-Min Wang}
   \affiliation{\ihep}
   \affiliation{\UCASastro}
   \affiliation{\NAOC}
\collaboration{22}{(SARM Collaboration)}

\begin{abstract}
We conduct a reverberation mapping (RM) campaign to spectroscopically monitor a sample of selected bright active galactic nuclei with large anticipated
broad-line region (BLR) sizes adequate for spectroastrometric observations by the GRAVITY instrument on
the Very Large Telescope Interferometer. We
report the first results for five objects, IC 4329A, Mrk 335, Mrk 509, Mrk 1239, and PDS 456, among which Mrk 1239 and PDS 456 are for the first time
spectroscopically monitored. We obtain multi-year monitoring data and perform multi-component spectral decomposition to extract the broad H$\beta$ profiles.
We detect significant time lags between the H$\beta$ and continuum variations, generally obeying the previously established BLR size-luminosity relation. Velocity-resolved H$\beta$ time lags illustrate diverse, possibly evolving BLR kinematics. We further measure the H$\beta$ line widths from mean and rms spectra and the resulting virial products show good consistency among different seasons. Adopting a unity virial factor and the full width at half maximum of the broad H$\beta$ line from the mean spectrum as the measure of velocity, the obtained black hole mass averaged over seasons is
$\log M_\bullet/M_\odot=8.02_{-0.14}^{+0.09}$, $6.92_{-0.12}^{+0.12}$, $8.01_{-0.25}^{+0.16}$, $7.44_{-0.14}^{+0.13}$, and $8.59_{-0.11}^{+0.07}$
for the five objects, respectively. The black hole mass estimations using other line width measures are also reported (up to the virial factors). For objects with previous RM campaigns, our mass estimates are in agreement with earlier results.
In a companion paper, we will employ BLR dynamical modeling to directly infer the black hole mass and thereby determine the virial factors.
\end{abstract}

\section{Introduction}
Supermassive black holes (SMBHs) are the dominant component of active galactic nuclei (AGNs). They provide a power engine for the huge AGN radiations through gaseous accretion from the surrounding medium. Measuring masses of SMBHs has remained challenging because at a cosmic distance, the quite compact spheres of gravitational influence of SMBHs are hardly resolved by existing telescopes, except for cases of very nearby quiescent galaxies (\citealt{Kormendy2013}).
Recently, a new joint analysis that combines spectroastrometry (SA) and reverberation mapping (RM) observations of AGNs (hereafter dubbed as SARM\footnote{The SARM analysis code \textsc{BRAINS} developed in \cite{Wang2020} and subsequent works (\citealt{Li2022, Li2023})  is open-source software and publicly available at \url{https://github.com/LiyrAstroph/BRAINS}.}) was proposed and applied by \cite{Wang2020} to simultaneously measure the SMBH mass and geometric cosmic distance of the quasar 3C~273. This SARM analysis makes full use of the complementary information about the broad-line emitting region in the AGN conveyed from SA and RM techniques and therefore can improve accuracy of the SMBH mass measurement (\citealt{Rakshit2015}).
The second application case of SARM analysis had been subsequently conducted upon NGC~3783 by \cite{Gravity2021}. These studies reinforce the great potential of SARM analysis for SMBH mass measurements as well as cosmic distance probes for cosmology \citep{Songsheng2021}.


It is known that AGN spectra are characterized by the presence of various broad emission lines with typical widths of
several thousands of kilometer per second, superimposed upon a broad-band continuum. While the continuum is believed to emanate from
the accretion disk surrounding the central SMBH, those broad emission lines come from an outer, but still compact region
in the center of the AGN, the so called broad-line region (BLR). This region
has a characteristic size ranging from several light-days to several hundreds of light-days, mainly depending on the AGN luminosity (e.g., \citealt{Kaspi2000, Bentz2013, Du2019}). Such a size is still well within the sphere of the SMBH's gravitational influence. The velocity distribution of BLR gas clouds is therefore dominated by the SMBH's gravity and encapsulates the information of the SMBH mass.

For a long period, the RM technique stood out as the uniquely practical way to probe BLR structure and kinematics and thereby estimate SMBH masses (\citealt{Bahcall1972, Blandford1982, Peterson2004}).
The situation changed with the operation of the GRAVITY instrument on the Very Large Telescope Interferometer (VLTI; \citealt{Gravity2017}), which allows to directly resolve BLR structure with the aid of the SA technique (e.g. \citealt{Petrov2001, Marconi2003,Gnerucci2010,Stern2015,Gravity2018,Bosco2021,Li2023}).

\subsection{Reverberation Mapping}
Specifically, the RM technique swaps spatial resolution for time resolution through capturing echoes of broad emission lines to the continuum variations in time domain (\citealt{Bahcall1972, Blandford1982}). These echoes show time delays, due to light travel time from the accretion disk to the BLR, as well as velocity dependence due to different Doppler shifting velocities from different parts of the BLR. By analyzing velocity-resolved echoes of broad emission lines, the modern dynamical modeling approach can directly determine the SMBH mass (\citealt{Pancoast2011, Pancoast2014, Li2013, Li2018}).

Separately, on the basis that the BLR motion is virialized (\citealt{Peterson1999}), a simple combination of the BLR size and
square of the line velocity widths constitutes a virial product that provides an approximate indicator to the SMBH mass. In practice,
an additional factor has to be invoked to convert the virial product to true SMBH mass, which needs to be calibrated otherwise.
There is evidence that the virial factor might vary from object to object (\citealt{Pancoast2014b, Grier2017, Li2018, Mejia2018, Williams2018, Villafana2022,Gravity2024}).
This implies that
the averaged value of the virial factor from the statistical calibration with the relationship between SMBH masses and stellar velocity
dispersions of the host galaxies (e.g., \citealt{Onken2004, Ho2014, Yu2019}) might introduce systematic errors to the estimated SMBH masses.
Nevertheless, we note that such a virial product-based SMBH mass measurement lays the foundation for large-scale SMBH demography. Therefore, a comprehensive study on how the virial factor depends on various AGN properties would be of central importance.

The information of BLR structure and kinematics delivered from RM indeed bears limitations in the sense that
RM only probes two dimensions of BLRs, namely, the time lag and line-of-sight (LOS) velocity. Mapping out the overall
BLR structure and kinematics (with six dimensions) from the observed time lags and LOS velocities is not straightforward and generally
some prior restrictions/presumptions on BLR models have to be involved. Any independent BLR diagnostics would be beneficial
to surmounting the limitations of RM and  advancing our understanding on BLR properties.

\subsection{Spectroastrometry}
As mentioned above, infrared interferometry
has seen great progress recently towards spatially resolving BLRs in bright AGNs based on the SA technique, marked by the successful
observations on the Pa$\alpha$ BLR of 3C 273 with the GRAVITY/VLTI instrument (\citealt{Gravity2018}). This supplies the sought-after new dimensions of BLRs and thereby provides an independent sort
of BLR diagnostics.

The SA technique measures astrometry (or photocenters) of the BLR as a function of wavelength/velocity. In addition to interferometry, SA can also be observed through spectroscopy with large single-aperture telescopes (\citealt{Stern2015, Bosco2021, Li2023}). The prominent photocenter offsets between red and blue shoulders of the broad Pa$\alpha$  line observed in 3C 273 (\citealt{Gravity2018}) clearly indicate coherent motion of BLR clouds (\citealt{Songsheng2019}).
Such an SA signal along with the broad line profile convey information of the BLR and therefore can be used to measure the black hole mass, provided with an appropriate BLR model.

Although current applications of SA observations on BLRs are still rare (e.g., \citealt{Gravity2018, Gravity2020, Gravity2021, Gravity2024, Abuter2024})
due to the rather strict magnitude limit ($K\lesssim10$), we expect that future SA observations will get significant growth
with the forthcoming upgraded instrument GRAVITY+ (\citealt{Gravity2022}), as well as once other ground- and space-based
interferometer arrays come into operation, such as the Magdalena Ridge Observatory Interferometer (MROI; \citealt{Creech2018})
and the Large Interferometer For Exoplanets (LIFE; \citealt{Quanz2022}).

\subsection{SARM Analysis}
The capability of SA and RM techniques are complementary in  two aspects. First, SA probes BLR structures perpendicular to the LOS, namely, the projected structures onto the observer's sky, whereas RM probes structures perpendicular to the iso-delay surfaces because of its resort to time delays between variations of the emission line and continuum. As a result, SARM analysis delivers more thorough, unique information of BLR kinematics and structures and thereby can improve SMBH measurements (e.g., \citealt{Wang2020, Li2022, Li2023}). It is clear that high-precision SMBH measurements are vital to exploring different evolution schemes of the co-evolution between SMBHs and their host galaxies (e.g., \citealt{Zhuang2023}).
Second, compared to RM that measures a linear size of the BLR, SA measures the angular size. 
SARM analysis naturally yields pure geometric distances of AGNs (\citealt{Wang2020, Li2023}). Such an AGN-based cosmic geometric probe is expected to be of significance for cosmology, considering the ever increasing tension between early and late universe measurements of the Hubble constant (e.g., \citealt{Riess2020}).
Indeed, the potential of the joint SA and RM analysis had been previously realized by \cite{Elvis2002} and \cite{Rakshit2015} from a theoretical aspect, but the practical application had not been feasible until the GRAVITY instrument came into operation. As mentioned above, the first application was conducted by \cite{Wang2020} upon 3C 273 and the second by \cite{Gravity2021} upon NGC\,3783. 

According to different modes of combination between SA and RM observations, SARM analysis can be divided into 1D, 2D, and time-domain SARM. To be specific, 1D SARM only uses the velocity-integrated line light curve from RM, and 2D SARM includes the velocity-resolved line RM data (\citealt{Li2022}). The time-domain SARM measures time variations of the SA signal and directly performs spectroastrometric reverberation mapping analysis (see \citealt{Li2023} for more details).

Given the above motivations and progress, we devised an SARM project to conduct both SA and RM experiments on a sample of selected AGNs with adequate BLR sizes suitable for SA observations
with GRAVITY/VLTI and other optical interferometers.
The first step would be to start an RM campaign to spectroscopically monitor those AGNs. As the first in a series paper, this work reports the RM results for five targets.
In a subsequent paper, we will analyze the obtained spectroscopic data with the BLR dynamical modeling approach and present the detailed BLR
structure and kinematics as well as direct black hole mass measurements for the five targets.

The paper is organized as follows. In Section 2, we present target selection criteria for the five targets and their observations and data reduction.
In Section 3, we describe detailed spectral decomposition to extract the H$\beta$ profiles and estimate the host galaxy fluxes. In Section 4,
we show the obtained H$\beta$ light curves and measure their time lags with respect to the photometric light curves.
In Section~5, we measures H$\beta$ line widths from the mean and root-mean-square (rms) spectra and combine them with H$\beta$ time lags
to estimate the black hole masses. Section 6 presents discussions on the obtained results, followed by conclusions in Section~7.

Throughout the paper, we use the standard $\Lambda$CDM cosmology with $H_0=67~{\rm km~s^{-1}~Mpc^{-1}}$, $\Omega_{\rm M}=0.32$, and $\Omega_{\Lambda}=0.68$
(\citealt{Planck2020}).


\begin{deluxetable*}{cccccccccccccc}
\tabletypesize{\footnotesize}
\tablecaption{Properties of Objects\label{tab_sample}}
\tablecolumns{13}
\tablewidth{0.95\textwidth}
\tablehead{
\colhead{~~~Name~~~} & \colhead{$z$} & \colhead{$D_L$} & $m_V$ & $m_K$ & \colhead{$A_V$}  & \colhead{$S_{5100}$}     & \colhead{$N_{\rm obs}$} & \colhead{$\Delta t_{\rm med}$} & \colhead{JD}  & \colhead{Year} & \colhead{$\Delta T$} & \colhead{~~~~~~Telescopes~~~~~~}  \\
\colhead{}     & \colhead{}  & \colhead{(Mpc)} & \colhead{(mag)} &  \colhead{(mag)}      & \colhead{(mag)}   & \colhead{}   & \colhead{}              & \colhead{(day)}  & \colhead{(-2,450,000)} & & \colhead{(day)}\\
\colhead{(1)}   & \colhead{(2)} & \colhead{(3)} & \colhead{(4)} & \colhead{(5)} & \colhead{(6)} & \colhead{(7)} & \colhead{(8)} &\colhead{(9)} &\colhead{(10)} &\colhead{(11)} & \colhead{(12)} & \colhead{(13)}
}
\startdata
IC 4329A & 0.0161 & 72.7 & 13.54 &  9.25 & 0.159 (+2.015) & 11.08 & 121  &  3 & 8,881-9,801  & 2020-2022 & 920 & LJ, CAHA, SAAO\\
Mrk 335  & 0.0258 & 117.6    & 14.33 & 10.54 & 0.096      & 1.10 & 149 & 4 & 8,693-9,981  & 2019-2023 & 1288 & CAHA\\
Mrk 509  & 0.0344 & 157.9    & 13.54 & 10.19 & 0.157      & 1.16 & 146 & 5 & 8,620-9,909  & 2019-2022 & 1289 & CAHA, SAAO\\
Mrk 1239 & 0.0199 & 90.5     & 14.27 &  9.69 & 0.175 (+1.674) & 6.60 & 106  & 5 & 8,775-9,735  & 2020-2022 & 959 & LJ, CAHA\\
PDS 456  & 0.1850 & 936.3    & 14.33 &  9.83 & 1.418 & 3.07 & 147  & 4 & 8,991-10,236  & 2018-2023 & 1245 & CAHA
\enddata
\tablecomments{Column (1): AGN name. Column (2): redshift taken from NASA/IPAC Extragalactic Database (NED).
Column (3): luminosity distance derived from redshift.
Columns (4) and (5): $V$ and $K$ band magnitudes taken from the compilation of \cite{Wang2020}. Note that the $K$ band magnitude represents the integrated nuclear magnitude with an aperture of 4\arcsec, thereby including contributions from the host galaxy. Column (6): the Galactic foreground $V$ band extinction
and the intrinsic extinction indicated by the number in parenthesis.
Column (7): the factor arising from the Galactic
foreground and intrinsic extinction in converting the observed flux density at $5100(1+z)$~{\AA} to the rest-frame flux density at 5100~{\AA}.
Column (8): number of observation epochs. Column (9): median sampling interval. Column (10): Julian days of observations.
Column (11): calendar years of observations. Column (12): observation period. Column (13): telescopes used for observations. ``LJ'' means Lijiang 2.4m telescope, ``CAHA'' means CAHA 2.2m telescope, and ``SAAO'' means SAAO 1.9m telescope. }
\end{deluxetable*}

\begin{figure*}[t]
\centering
\includegraphics[width=0.85\textwidth]{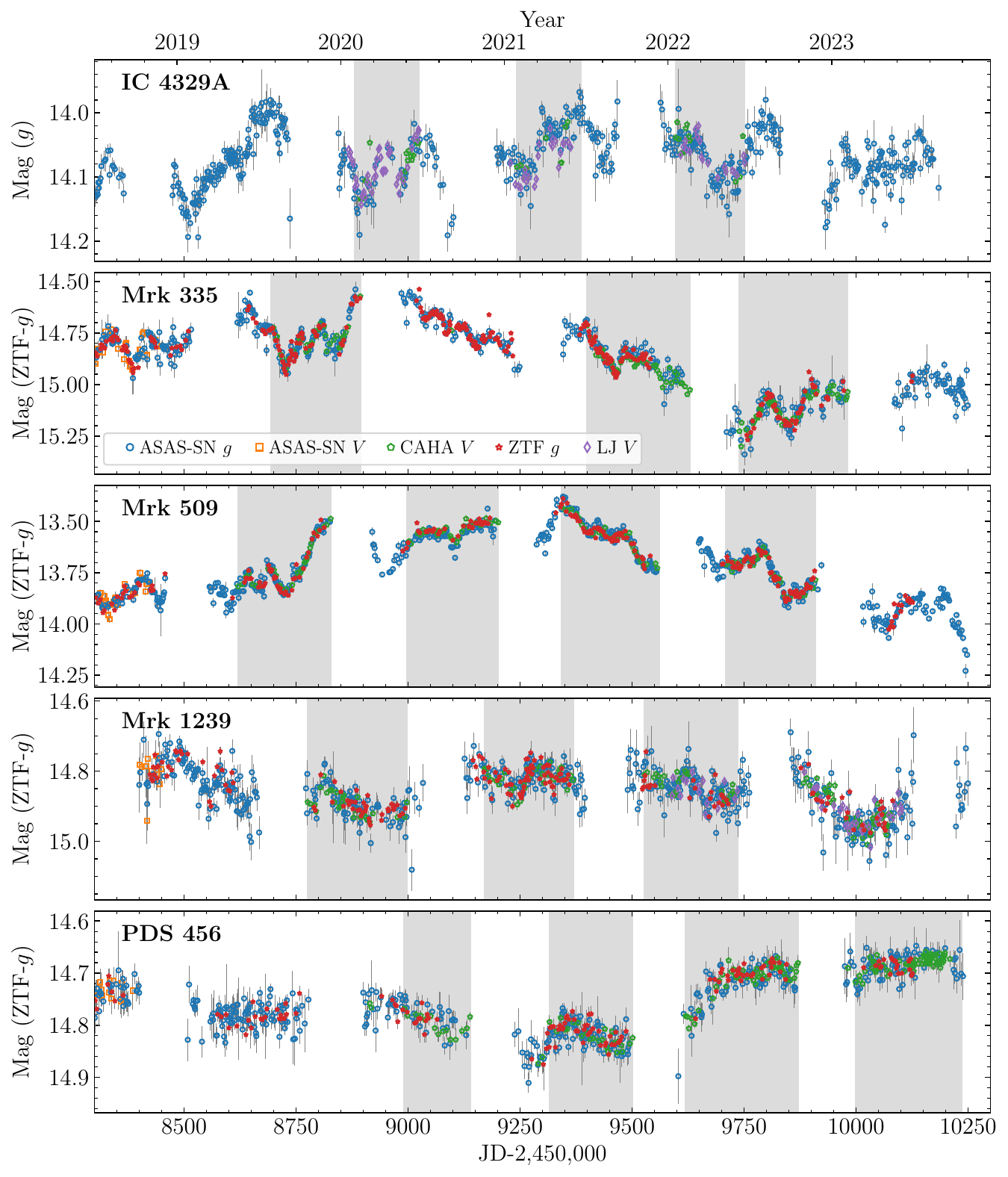}
\caption{Intercalibrated photometric light curves. The reference dataset of intercalibration is by default adopted to be ZTF-$g$ band photometry for
Mrk 335, Mrk 509, Mrk 1239, and PDS 456. For IC 4329A, there is no available ZTF data and the ASAS-SN $g$ band photometry is chosen as the reference dataset.
Gray shaded bands represent the spectral monitoring periods.}
\label{fig_con_lc}
\end{figure*}

\begin{figure*}
\centering
\includegraphics[width=0.9\textwidth]{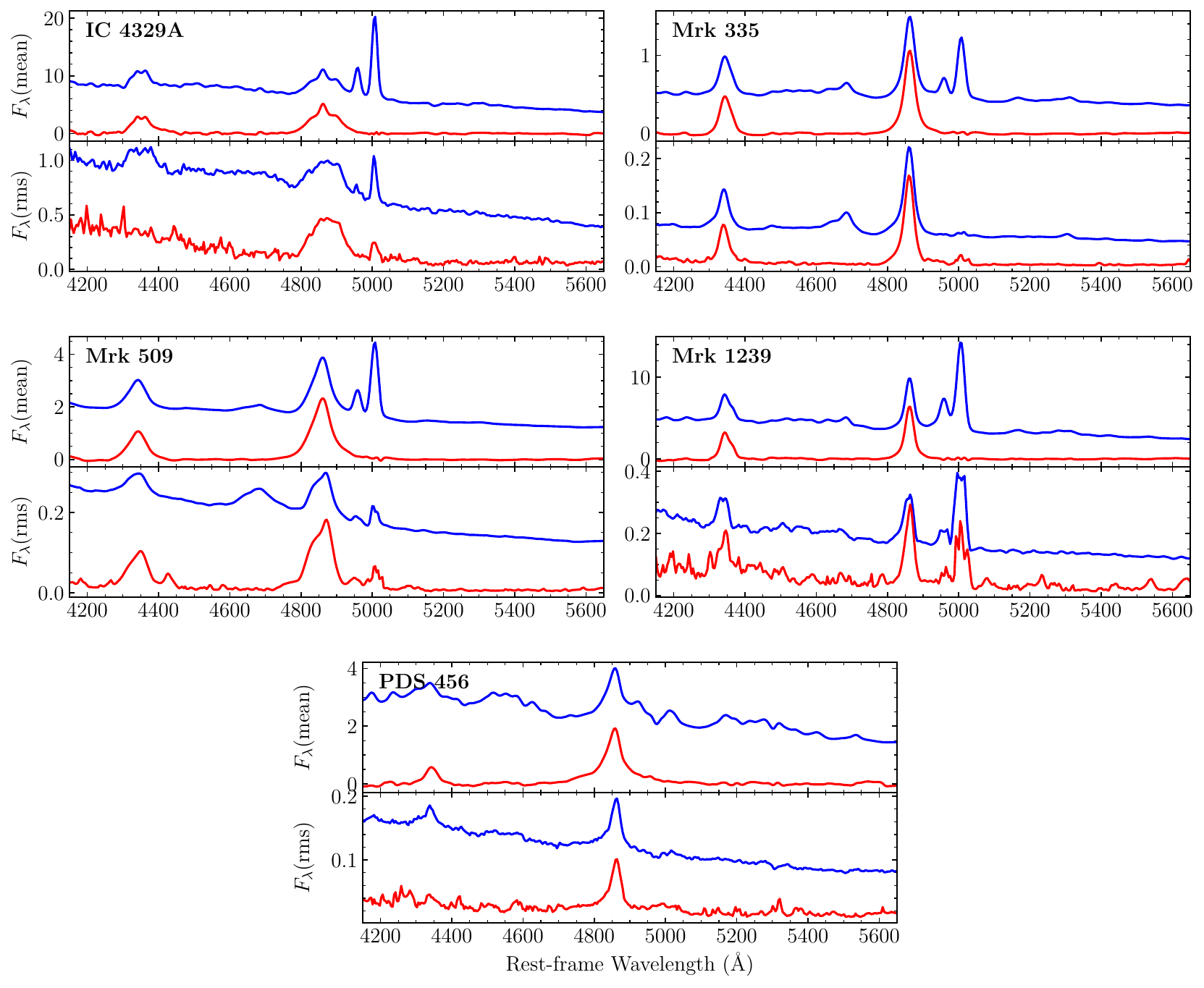}
\caption{The mean and rms spectra over the whole monitoring period. For each object, the top and bottom panels show the mean and rms spectra, respectively. The red lines correspond
to spectra after subtracting other spectral components except for the H$\beta$ and H$\gamma$ lines. The blue lines corresponds to spectra without subtraction of
other spectral components.
All spectra are in units of $10^{-14}~\rm erg~s^{-1}~cm^{-2}~\text{\AA}^{-1}$.
{\it Note that all spectra have been corrected for the Galactic foreground and intrinsic extinction.}}
\label{fig_mean_rms}
\end{figure*}

\section{Observations and Data Reduction}
\subsection{Targets}\label{sec_targets}
We select five targets, IC4329A, Mrk 335, Mrk 509, Mrk 1239, and PDS 456, from a sample of infrared-bright AGNs compiled by \cite{Wang2020}, which are based on existing known AGN catalogues with the criteria of declination $<20^\circ$ and $K$-band magnitude $K<11.5$.
These AGNs are possible targets for GRAVITY or future GRAVITY+ interferometric observations. All the five targets have $K<10.5$
and anticipated BLR angular sizes $\gtrsim20$~microarcseconds (without including the line strength and position angle; see Table 1 of the Supplementary Information in \citealt{Wang2020}). 
Note that here, the magnitudes are taken from the Two Micron All Sky Survey, which refer to the integrated magnitudes within an aperture of 4{\arcsec}. As a result, the magnitudes encompass the contributions from the host galaxy and thereby overestimate the nuclear magnitudes. Nevertheless, the five targets had been observed by GRAVITY/VLTI and their dusty torus sizes in $K$ band had been estimated based on the interferometric visibility observations (priviate communication with E. Sturm; see also \citealt{Gravity2023}). Recently, the BLR sizes of IC~4329A, Mrk 335, Mrk 509, and PDS 456 had also been estimated based on the differential phase observations (\citealt{Gravity2024}).

Regarding RM observations, Mrk 1239 and PDS 456 are spectroscopically monitored for the first time. The other three objects have previous
RM measurements by different RM campaigns (Mrk 509: \citealt{Peterson1998a};
Mrk 335: \citealt{Peterson1998a,Grier2012, Du2014, Hu2015, Hu2021}; IC 4329A: \citealt{Bentz2023}).
Here, we provide new multi-year spectroscopic monitoring results. The basic properties of the five objects are summarized in Table~\ref{tab_sample}.

\subsection{Photometry}\label{sec_photometry}
Our spectral and photometric observations were mainly undertaken with the Lijiang (LJ) 2.4m telescope at the Yunnan Observatory
of the Chinese Academy of Sciences and the Centro Astron\'omico Hispano-Alem\'an (CAHA) 2.2m telescope.
We took three broad-band photometric images of the targets with a Johnson $V$ filter, just prior to the spectral exposures.
The typical image exposure time was 10-20s each.

The data are reduced using the standard IRAF procedures, as usual including bias removal, flat-fielding, and then aperture photometry.
We choose 3-6 nearby reference stars around the target and measure the differential photometry of the target with respect to these reference stars.
The associated errors are assigned by the Poisson noises in photon counts of both the target and reference stars.
Here, the absolute flux calibration is not important because the obtained differential photometry will be aligned with other datasets below.

In addition to our own photometric observations, we also collect archival photometric data from two public time-domain surveys, namely,
the All-Sky Automated Survey for Supernovae (ASAS-SN; \citealt{Shappee2014, Kochanek2017}) and the Zwicky Transient Facility (ZTF; \citealt{Graham2019}).
This significantly improves photometric sampling cadences.
ASAS-SN provides $V$ band photometry from late 2012 to mid 2018 and $g$ band photometry from late 2017 till now. ZTF provides three-band (ZTF-$g, r, i$) photometry since 2018.
In the following analysis, we use both $g$ and $V$ band photometry of ASAS-SN and $g$ band photometry of ZTF.

We intercalibrate photometry from these different sources using the Bayesian package {\textsc{PyCALI}}\footnote{The living code is publicly available at \url{https://github.com/LiyrAstroph/PyCALI}, while the version used in this work is available at \url{https://doi.org/10.5281/zenodo.10700132}.}
(\citealt{Li2014}). {\textsc{PyCALI}} employs a multiplicative factor $\varphi$ and an additive factor $G$ to shift and scale a dataset to align with
the reference dataset as (e.g, \citealt{Peterson1995})
\begin{equation}
F_{\rm cali} = \varphi F_{\rm obs} - G,
\end{equation}
where $F_{\rm obs}$ is the original flux densities and $F_{\rm cali}$ is the intercalibrated flux densities.
The factors $\varphi$ and $G$ are different for different datasets and for the reference dataset, $\varphi=1$ and $G=0$.
We first convert magnitudes into flux densities with an arbitrary zero-point flux density. After
intercalibration, we convert the flux densities back to magnitudes.
As such, we effectively intercalibrate the magnitudes.
{\textsc{PyCALI}} assumes that the light curve variability follows a damped random walk process and
determines the best estimates of $\varphi$ and $G$ for each data source using a Markov-chain Monte Carlo method.
The uncertainties and covariances of $\varphi$ and $G$ are also incorporated into the uncertainties of final intercalibrated magnitudes.
It is worth mentioning that through simulation tests, we verify that the obtained factors are largely independent on
the intrinsic variability model of the light curve.
We by default adopt the ZTF-$g$ band photometry (with a center wavelength of 4806~{\AA}) as the reference dataset for Mrk 335, Mrk 509, Mrk 1239, and PDS~456.
However, for IC 4329A, there is not available ZTF data. We instead use $g$ band photometry of ASAS-SN (with a center wavelength
of 4747~{\AA}) as the reference dataset.

There are a few apparent outlier points in the ASAS-SN and ZTF data. We use the reconstruction to the combined light curve output from \textsc{PyCALI} based on the damped random walk process and identify those points as outliers with their deviations from the reconstruction larger than 5$\sigma$.
We then discard these outliers and rerun the above intercalibration procedure again to obtain the final combined photometric light curves.
Figure~\ref{fig_con_lc} shows the final light curves for five objects. As can be seen, {\textsc{PyCALI}} almost entirely removes
systematic differences between different datasets and well aligns the variation patterns within uncertainties.

\begin{figure*}
\centering
\includegraphics[width=0.8\textwidth]{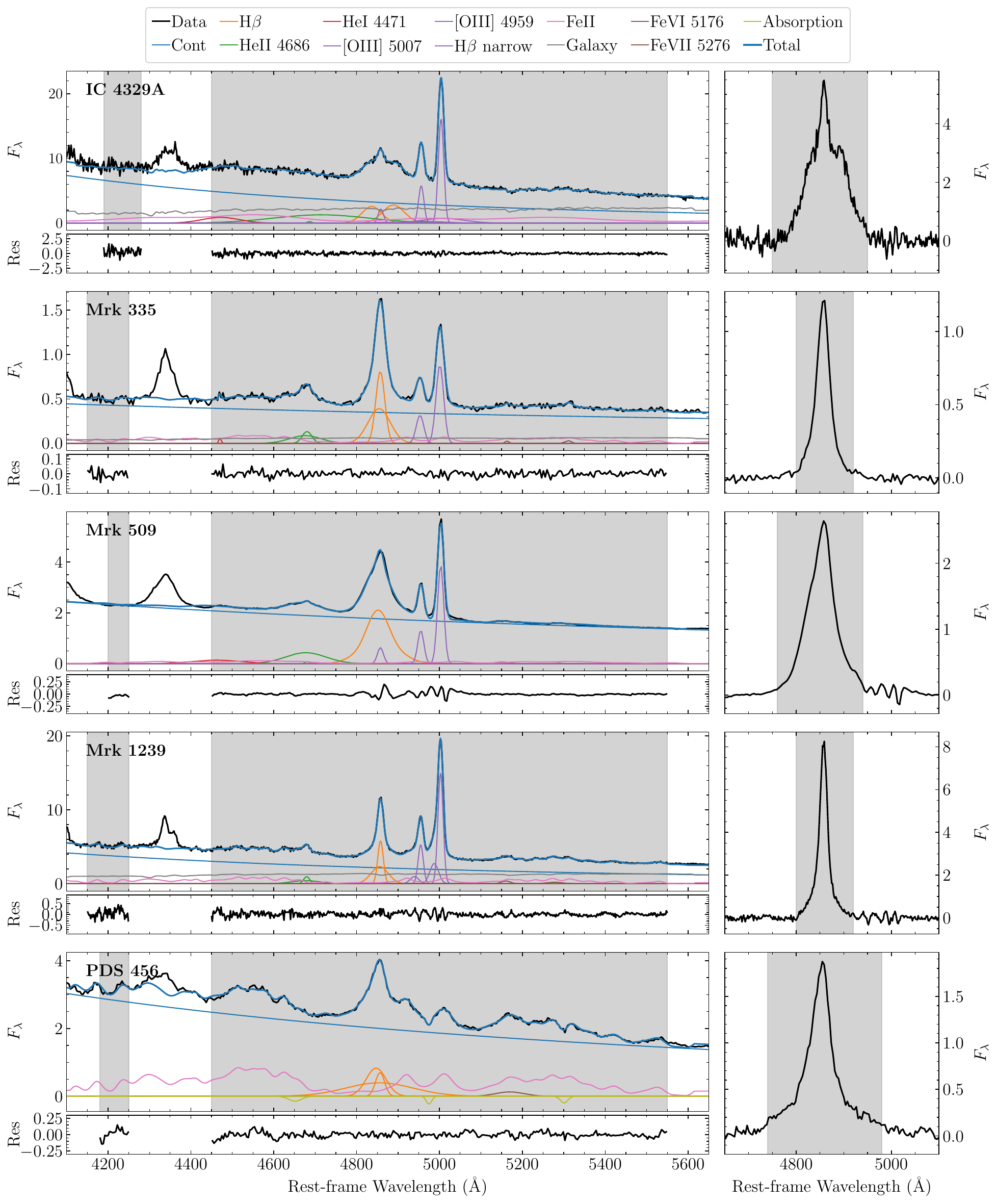}
\caption{Examples of the spectral decomposition. For each object, the left panels show the spectral fitting with various components and the residuals.
Gray shaded bands represent the fitting windows. The right panel show the extracted H$\beta$ line and the spectral window adopted for calculating
integrated H$\beta$ flux. The flux densities $F_\lambda$ are all in units of
$10^{-14}~\rm erg~s^{-1}~cm^{-2}~\text{\AA}^{-1}$. {\it Note that all spectra have been corrected for the Galactic foreground and intrinsic extinction.}}
\label{fig_specfit}
\end{figure*}

\begin{figure*}
\centering
\includegraphics[width=0.85\textwidth]{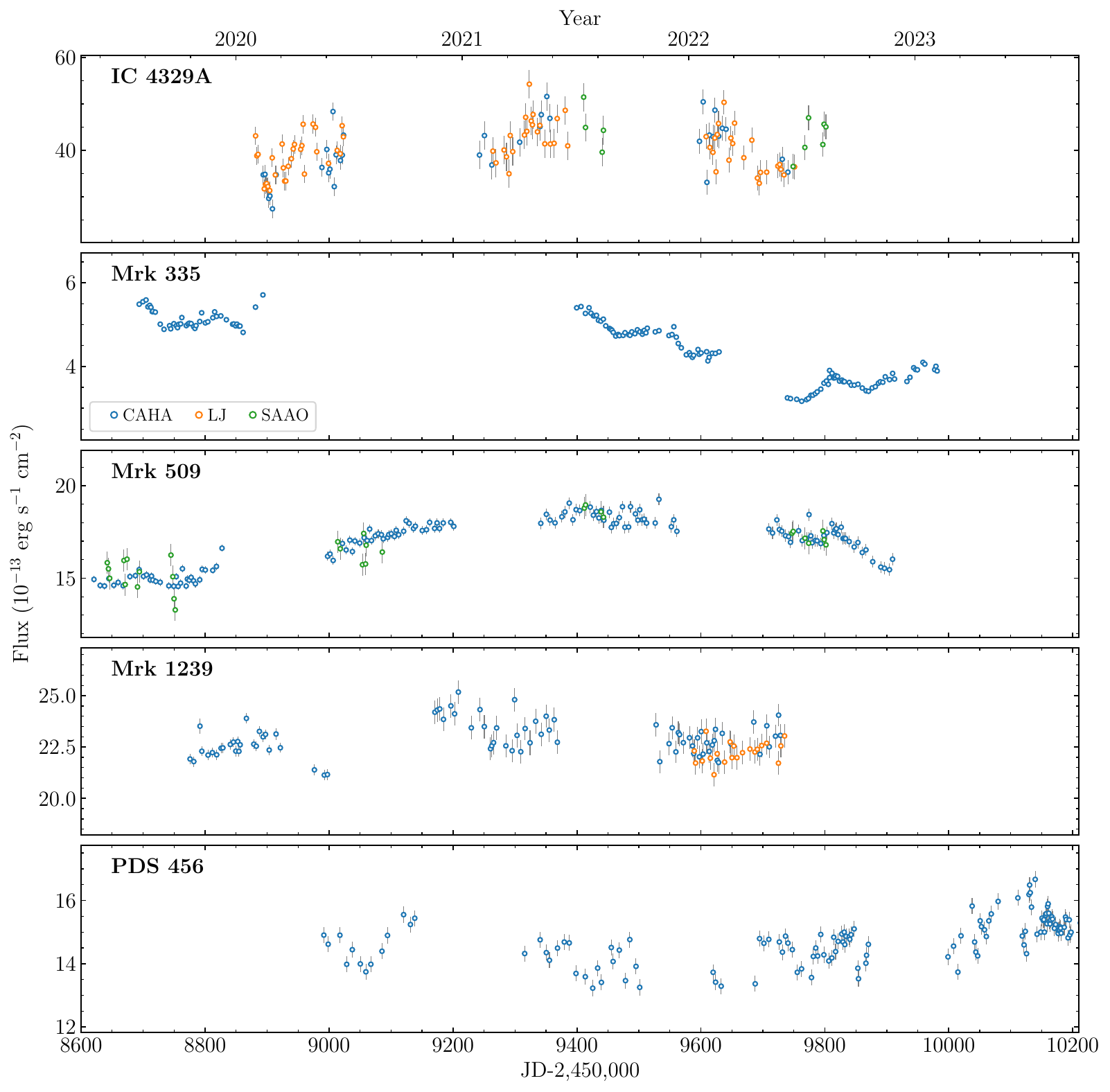}
\caption{The light curves of the integrated H$\beta$ fluxes over the whole monitoring period. }
\label{fig_hb_lc}
\end{figure*}

\subsection{Spectroscopy}
The majority of spectra were taken with the Calar Alto Faint Object Spectrograph on CAHA 2.2m telescope and the Yunnan Faint Object Spectrograph and Camera on Lijiang 2.4m telescope.  A small fraction of spectra were also taken with the SpUpNIC Spectrograph on the 1.9m telescope at the Sutherland station of the South African Astronomical Observatory (SAAO). The observation procedures are similar except that different grims and slit widths were used on the three telescopes in consideration of different weather conditions.

For CAHA observations, we used the Grim G-200 and a long slit with a width of 3{\arcsec}. This gives a wavelength cover range of
4000-8500 \AA\ and a dispersion of 4.47 \AA\ pixel$^{-1}$.
For Lijiang observations, we used the Grism 14 and a long slit with a width of 2.5{\arcsec}, resulting in a wavelength cover range
of 3600-7460 \AA\ and a dispersion of 1.8 \AA\ pixel$^{-1}$.
For SAAO observations, we used the grating 6 and a slit with a width of 4.04{\arcsec}, which gives a wavelength range of $\sim$3600-6400 {\AA} and a dispersion of {1.36 {\AA} pixel$^{-1}$}. 


During CAHA and LJ observations, the long slit was rotated to simultaneously observe a nearby comparison star during the spectral exposure
of the target (see \citealt{Maoz1990}). This comparison star is used for the following two-step flux calibration.  Firstly, we generate the fiducial spectrum of the comparison star using the observed data of spectrophotometric standards on photometric nights. Secondly, we then derive a wavelength-dependent sensitivity function by comparing the extracted spectrum of the star to its fiducial spectrum. This sensitivity function is subsequently applied to the extracted spectrum of the target to obtain its calibrated spectrum (see more details in \citealt{Du2014} and \citealt{Hu2021}). Our previous observations had well demonstrated that such a calibration strategy
can typically yield a flux precision better than $3\%$ (\citealt{Du2014, Du2016, Du2018, Hu2020a, Hu2020b, Lu2022}). For SAAO observations, because the slit cannot be rotated to accommodate a comparison star, the spectra were taken in a normal way and calibrated using a standard star.

Two exposures of 600s each were taken for each object in one night. The spectral reduction follows the standard IRAF procedures for bias, flat-fielding, and wavelength calibrations. The spectral extraction apertures are adopted to be in a range of 8.5{\arcsec}-16.3{\arcsec}, adjusted accordingly for different telescopes. The larger extraction aperture than the slit width was used to minimize seeing-dependent light loss.

After spectral reduction, we correct all spectra for the Galactic foreground extinction using the $R_V$-dependent Galactic extinction law of \cite{Fitzpatrick2004} with $R_V=3.1$.
The $V$-band extinction $A_V$ is taken from \cite{Schlafly2011} and listed in Table~\ref{tab_sample}. For IC 4329A and Mrk 1239, previous studies showed that they both have a heavy intrinsic
extinction (\citealt{Marziani1992,Mehdipour2018, Rodriguez-Ardila2006, Pan2021}), which is also supported by their red spectral shapes in our observations. As a comparison, the rest three objects (Mrk 335, Mrk 509, and PDS 456) display blue spectral shapes and an extra intrinsic extinction is thus not included.
The intrinsic extinction of IC 4329A
is set by a color excess $E(B-V)=1.0$ following \cite{Mehdipour2018}, which corresponds to a $V$-band extinction $A_V=2.015$
using the extinction law of \cite{Czerny2004}. The intrinsic extinction of Mrk 1239 is set by a color excess $E(B-V)=0.54$
following \cite{Rodriguez-Ardila2006}, which corresponds to a $V$-band extinction $A_V=1.674$ using the extinction law of the Small
Magellanic Cloud (\citealt{Gordon2003}). 
The adopted magnitudes of $V$-band intrinsic extinction for IC~4329A and Mrk 1239 are also listed in Table~\ref{tab_sample}.

In Figure~\ref{fig_mean_rms}, we show mean and root-mean-square (rms) spectra for all five objects over the whole monitoring period.
The mean and rms spectra are calculated using the signal-to-noise ratio (S/N) as a weight following \cite{Park2012}, namely,
\begin{gather}
\bar{f}(\lambda) = \sum_{i=1}^{N} w_i f_i(\lambda),\label{eqn_mean}\\
\sigma(\lambda)  = \sqrt{\frac{1}{1-\sum_{i=1}^{N}w_i^2}\sum_{i=1}^{N}w_i\left[f_i(\lambda)-\bar{f}(\lambda)\right]^2},\label{eqn_rms}
\end{gather}
where $N$ is the number of spectra and the weight is
\begin{equation}
w_i(\lambda) = \frac{({\rm S/N})_i(\lambda)}{\sum_{i=1}^{N}({\rm S/N})_i(\lambda)}.
\end{equation}
For the sake of comparison, also superimposed in Figure~\ref{fig_mean_rms} are
the mean and rms spectra after subtracting other emission components but for H$\beta$ and H$\gamma$ lines
based on the spectral decomposition procedure described in next section.

\begin{deluxetable*}{cccccccccccccc}
\tabletypesize{\footnotesize}
\tablecaption{Light Curve Properties\label{tab_lc}}
\tablecolumns{14}
\tablewidth{0.95\textwidth}
\tablehead{
\colhead{Name} & Season &  \multicolumn{2}{c}{Photometry}  & & \multicolumn{5}{c}{5100~{\AA}} & & \multicolumn{3}{c}{H$\beta$}\\\cline{3-4}\cline{6-10}\cline{12-14}
\colhead{}     &        & $\langle \rm mag \rangle$ & $F_{\rm var}$  && \colhead{$\langle F \rangle$\tablenotemark{\scriptsize a}} & \colhead{$\sigma_{\rm sys}$} & $\langle F_{\rm gal}\rangle$\tablenotemark{\scriptsize b} & $\langle F_{\rm iron}\rangle$\tablenotemark{\scriptsize b} & $F_{\rm var}$\tablenotemark{\scriptsize c}  && $\langle F_{\rm H\beta} \rangle$ & $\sigma_{\rm sys}$ & $F_{\rm var}$\tablenotemark{\scriptsize c}\\\cline{6-9}\cline{12-14}
\colhead{}     &        &     &\colhead{(\%)}  && \multicolumn{4}{c}{($10^{-15}\rm~erg~s^{-1}~cm^{-2}~{\text{\AA}^{-1}}$)} & (\%)    && \multicolumn{2}{c}{($10^{-13}\rm~erg~s^{-1}~cm^{-2}$)} & (\%)
}
\startdata
 IC 4329A &  2020 & $14.08\pm0.03$ & 2.6 && $49.15\pm5.34$ & 2.60 &           &          & 10.9 && $36.27\pm4.53$ & 1.97 & 12.5\\
         &  2021 & $14.06\pm0.03$ & 3.1 && $54.77\pm5.19$ & 1.56 &           &          & 9.5 && $43.47\pm4.78$ & 3.08 & 11.0\\
         &  2022 & $14.06\pm0.03$ & 2.7 && $51.73\pm4.85$ & 3.83 &           &          & 9.4 && $39.61\pm4.56$ & 2.79 & 11.5\\
         &   All & $14.07\pm0.03$ & 3.0 && $51.30\pm5.46$ & 2.68 & $19.90\pm3.13$ & $6.23\pm1.65$ & 10.6 && $38.99\pm5.25$ & 2.69 & 13.5\\
  Mrk 335 &  2019 & $14.78\pm0.07$ & 6.6 && $4.56\pm0.32$ & 0.05 &           &          & 7.0 && $5.13\pm0.22$ & 0.05 & 4.2\\
         &  2021 & $14.88\pm0.06$ & 5.2 && $4.11\pm0.24$ & 0.09 &           &          & 5.8 && $4.76\pm0.33$ & 0.05 & 7.0\\
         &  2022 & $15.11\pm0.07$ & 6.1 && $3.39\pm0.20$ & 0.05 &           &          & 6.0 && $3.62\pm0.25$ & 0.04 & 6.8\\
         &   All & $14.88\pm0.15$ & 14.0 && $4.00\pm0.54$ & 0.07 & $0.54\pm0.19$ & $0.21\pm0.03$ & 13.6 && $4.49\pm0.69$ & 0.05 & 15.4\\
  Mrk 509 &  2019 & $13.72\pm0.11$ & 9.8 && $13.44\pm1.46$ & 0.16 &           &          & 10.9 && $14.95\pm0.41$ & 0.19 & 2.7\\
         &  2020 & $13.55\pm0.05$ & 4.3 && $15.78\pm0.50$ & 0.16 &           &          & 3.1 && $17.36\pm0.46$ & 0.23 & 2.7\\
         &  2021 & $13.57\pm0.08$ & 7.0 && $15.08\pm1.21$ & 0.30 &           &          & 8.0 && $18.29\pm0.41$ & 0.32 & 2.2\\
         &  2022 & $13.74\pm0.08$ & 7.2 && $12.90\pm0.87$ & 0.21 &           &          & 6.7 && $17.00\pm0.70$ & 0.33 & 4.1\\
         &   All & $13.69\pm0.15$ & 13.8 && $14.21\pm1.58$ & 0.22 & \nodata      & $0.27\pm0.04$ & 11.2 && $16.86\pm1.32$ & 0.27 & 7.8\\
 Mrk 1239 &  2019 & $14.89\pm0.03$ & 2.6 && $29.77\pm1.23$ & 0.66 &           &          & 4.1 && $22.47\pm0.58$ & 0.23 & 2.5\\
         &  2020 & $14.82\pm0.03$ & 2.6 && $31.77\pm0.88$ & 0.71 &           &          & 2.8 && $23.45\pm0.87$ & 0.58 & 3.7\\
         &  2021 & $14.85\pm0.03$ & 3.0 && $31.17\pm1.49$ & 1.04 &           &          & 4.8 && $22.62\pm0.58$ & 0.54 & 2.5\\
         &   All & $14.87\pm0.06$ & 5.4 && $30.84\pm1.50$ & 1.00 & $9.36\pm4.09$ & $2.11\pm0.24$ & 4.9 && $22.79\pm0.78$ & 0.56 & 3.4\\
  PDS 456 &   All & $14.73\pm0.05$ & 4.9 && $19.69\pm1.03$ & 0.21 & \nodata       & $2.24\pm0.10$ & 5.2 && $14.75\pm0.68$ & 0.25 & 4.6\\
\enddata
\tablenotetext{a}{$\langle F\rangle$ is the averaged flux density at 5100~{\AA}, which includes both contributions from the host galaxy and \ion{Fe}{2} emission.}
\tablenotetext{b}{$\langle F_{\rm gal}\rangle$ and $\langle F_{\rm iron}\rangle$ are the averaged flux densities of the host galaxy and \ion{Fe}{2} emissions at 5100~{\AA}, respectively, which are estimated from spectral decomposition.}
\tablenotetext{c}{$F_{\rm var}$ is calculated using Equation~(\ref{eqn_fvar}) without including the systematic errors ($\sigma_{\rm sys}$).}
\end{deluxetable*}

\section{Spectral Decomposition}\label{sec_decomp}
To extract the broad H$\beta$ line, we perform spectral decomposition by fitting the spectra with a series of spectral components (e.g.,
see \citealt{Hu2008, Hu2015, Barth2015}). The generic spectral components include the following:
(1) a featureless single power-law continuum; (2) the optical \ion{Fe}{2} template from \cite{Boroson1992}; (3) the host galaxy template,
chosen to be a single stellar population model from \cite{Bruzual2003};
(4) the narrow emission lines [\ion{O}{3}] $\lambda\lambda$4959, 5007, \ion{He}{2} $\lambda$4686, and H$\beta$; (5) the broad emission lines
H$\beta$ and \ion{He}{2} $\lambda$4686. In realistic implementation, the adopted spectral components are slightly different from object
to object according to the features in the spectra. The fitting windows are chosen to be wide enough to appropriately constrain
the slope of the power-law continuum component. By default, we use the fiducial window of 4450-5550 {\AA} plus an extra narrow window around
4200 {\AA}. The width of the narrow window is adjusted for each object to avoid the H$\gamma$ and H$\delta$ lines.

We employ a Markov-chain Monte Carlo technique with
the diffusive nested sampling algorithm\footnote{We use our own developed
code {\textsc{CDNest}} for the diffusive nested sampling algorithm. The living code is publicly available at \url{https://github.com/LiyrAstroph/CDNest},
while the version used in this work is available at \url{https://doi.org/10.5281/zenodo.7809591}.} (\citealt{Brewer2011})
to optimize the spectral fitting and determine the best-fit parameters.
The best-fit velocity shifts of [\ion{O}{3}] $\lambda$5007, when applicable, are used to align the spectra in wavelength
for objects with strong [\ion{O}{3}] emissions. The exception is PDS~456, which does not show [\ion{O}{3}] emission.
We instead use the O$_2$ B-band atmospheric absorption trough around 6880~{\AA}, clearly seen in PDS~456 as all its observations are at relatively high airmass due to its southern declination
(see below for details).
In Figure~\ref{fig_specfit}, we show an example of spectral decomposition to a selected epoch for each object, along with
the extracted H$\beta$ line. The spectral windows adopted to calculate integrated H$\beta$ fluxes are also illustrated in
the right panels. The obtained H$\beta$ light curves over the whole monitoring period are plotted in Figure~\ref{fig_hb_lc}.

\begin{deluxetable*}{cccccccccc}
\tabletypesize{\footnotesize}
\tablecaption{Time Lag Measurements.\label{tab_lag}}
\tablecolumns{10}
\tablewidth{0.95\textwidth}
\tablehead{
\colhead{~~~~~~~~~Name~~~~~~~~} & ~~~~~~Season~~~~~~  & \multicolumn{2}{c}{ICCF} & & \multicolumn{2}{c}{MICA} & &\multicolumn{2}{c}{Angular Size}\\\cline{3-4}\cline{6-7}\cline{9-10}
\colhead{}     &        &~~~~~~$r_{\rm max}$~~~~~~ & ~~~~~~~~$\tau_{\rm cent}$~~~~~~~~ & & ~~~~~~~~~~$\tau_{\rm mica}$~~~~~~~~~~ & ~~~~~~~~~~$w_{\rm mica}$~~~~~~~~~~ & & \colhead{~~~~~~~~~$\Delta \theta$~~~~~~~~~} & \colhead{~~~~~~~~~$\Delta \theta_{\rm obs}$~~~~~~~~~}\\
\colhead{}     &        & & \colhead{(day)}  &  & \colhead{(day)} & \colhead{(day)} && \colhead{($\mu$as)} & \colhead{($\mu$as)}
}
\startdata
 IC 4329A &  2020 & 0.75 & $12.33_{-2.64}^{+1.44}$ && $12.71_{-2.02}^{+0.96}$ & $0.97_{-0.60}^{+1.70}$ && $29.86_{-6.39}^{+3.49}$  & $1.69_{-0.36}^{+0.20}$ \\
         &  2021 & 0.76 & $20.86_{-9.65}^{+2.53}$ && $22.39_{-3.34}^{+1.34}$ & $1.47_{-1.02}^{+3.61}$ && $50.52_{-23.37}^{+6.13}$  & $2.86_{-1.32}^{+0.35}$ \\
         &  2022 & 0.69 & $10.08_{-5.09}^{+9.70}$ && $14.85_{-2.38}^{+1.10}$ & $0.79_{-0.46}^{+2.70}$ && $24.41_{-12.33}^{+23.49}$  & $1.38_{-0.70}^{+1.33}$ \\
  Mrk 335 &  2019 & 0.76 & $14.86_{-2.15}^{+2.59}$ && $19.36_{-1.35}^{+1.27}$ & $11.96_{-1.58}^{+1.54}$ && $22.46_{-3.25}^{+3.91}$  & $1.27_{-0.18}^{+0.22}$ \\
         &  2021 & 0.90 & $16.31_{-2.41}^{+3.47}$ && $12.67_{-1.30}^{+1.17}$ & $15.25_{-1.60}^{+1.39}$ && $24.65_{-3.64}^{+5.24}$  & $1.40_{-0.21}^{+0.30}$ \\
         &  2022 & 0.85 & $12.09_{-2.43}^{+2.32}$ && $13.76_{-1.80}^{+1.63}$ & $12.82_{-3.48}^{+2.29}$ && $18.27_{-3.67}^{+3.51}$  & $1.03_{-0.21}^{+0.20}$ \\
  Mrk 509 &  2019 & 0.75 & $24.11_{-9.13}^{+5.40}$ && $39.27_{-6.23}^{+4.71}$ & $16.35_{-3.80}^{+4.22}$ && $27.37_{-10.36}^{+6.13}$  & $1.55_{-0.59}^{+0.35}$ \\
         &  2020 & 0.81 & \nodata && \nodata & \nodata && \nodata & \nodata\\
         &  2021 & 0.52 & $52.18_{-9.89}^{+6.61}$ && $52.12_{-9.64}^{+7.01}$ & $2.31_{-1.88}^{+10.26}$ && $59.23_{-11.23}^{+7.50}$  & $3.35_{-0.64}^{+0.42}$ \\
         &  2022 & 0.81 & $34.43_{-6.79}^{+3.88}$ && $39.81_{-4.36}^{+4.85}$ & $28.78_{-4.89}^{+3.71}$ && $39.08_{-7.71}^{+4.40}$  & $2.21_{-0.44}^{+0.25}$ \\
 Mrk 1239 &  2019 & 0.75 & $78.73_{-7.85}^{+7.20}$ && $84.16_{-11.21}^{+14.66}$ & $43.62_{-17.55}^{+22.99}$ && $153.73_{-15.33}^{+14.06}$  & $8.70_{-0.87}^{+0.80}$ \\
         &  2020 & 0.42 & $54.91_{-18.42}^{+24.75}$ && $51.16_{-11.53}^{+15.01}$ & $41.45_{-16.05}^{+10.86}$ && $107.22_{-35.97}^{+48.33}$  & $6.07_{-2.04}^{+2.74}$ \\
         &  2021 & 0.16 & \nodata && \nodata & \nodata && \nodata & \nodata\\
  PDS 456 &   All & 0.70 & $281.22_{-56.13}^{+49.29}$ && $251.68_{-45.61}^{+68.28}$ & $118.34_{-114.82}^{+85.68}$  && $61.67_{-12.31}^{+10.81}$  & $23.13_{-4.62}^{+4.05}$ \\
\enddata
\tablecomments{All time lags are given in the observed frame. $\tau_{\rm mica}$ and $w_{\rm mica}$ refer to the center and standard deviation of the Gaussian transfer function,
respectively. $\Delta \theta$ and $\Delta \theta_{\rm obs}$ refer to the maximum angular photocenter offset of the BLR estimated using $\tau_{\rm cent}$ without and with the inclusion of line strength ratio in Equations~(\ref{eqn_ang}) and (\ref{eqn_ang_obs}), respectively. Note that the geometry and position angle of the BLR are not included in estimating $\Delta \theta$ and $\Delta \theta_{\rm obs}$, therefore, the actual values might be even smaller. }
\end{deluxetable*}

After spectral decomposition, we subtract all model components other than H$\beta$ (including both broad and narrow components)
from each nightly spectrum and leave the H$\beta$ profiles as residuals. We then calculate mean and rms spectra from this time series of
residuals using Equations (\ref{eqn_mean}) and (\ref{eqn_rms}). Figure~\ref{fig_mean_rms} plots
these component-subtracted mean and rms spectra for comparison with the unsubtracted spectra.
We note that rms spectra usually have a non-zero continuum levels arising from photon-counting noises and residual noises in
spectral decomposition (e.g., see also \citealt{Barth2015}), while the continuum levels in mean spectra stay at zero as expected. The H$\gamma$ line variability of IC 4329A and PDS 456 is also overwhelmed by these noises and thus invisible in the component-subtracted rms spectra.

\begin{figure*}
\centering
\includegraphics[width=0.8\textwidth]{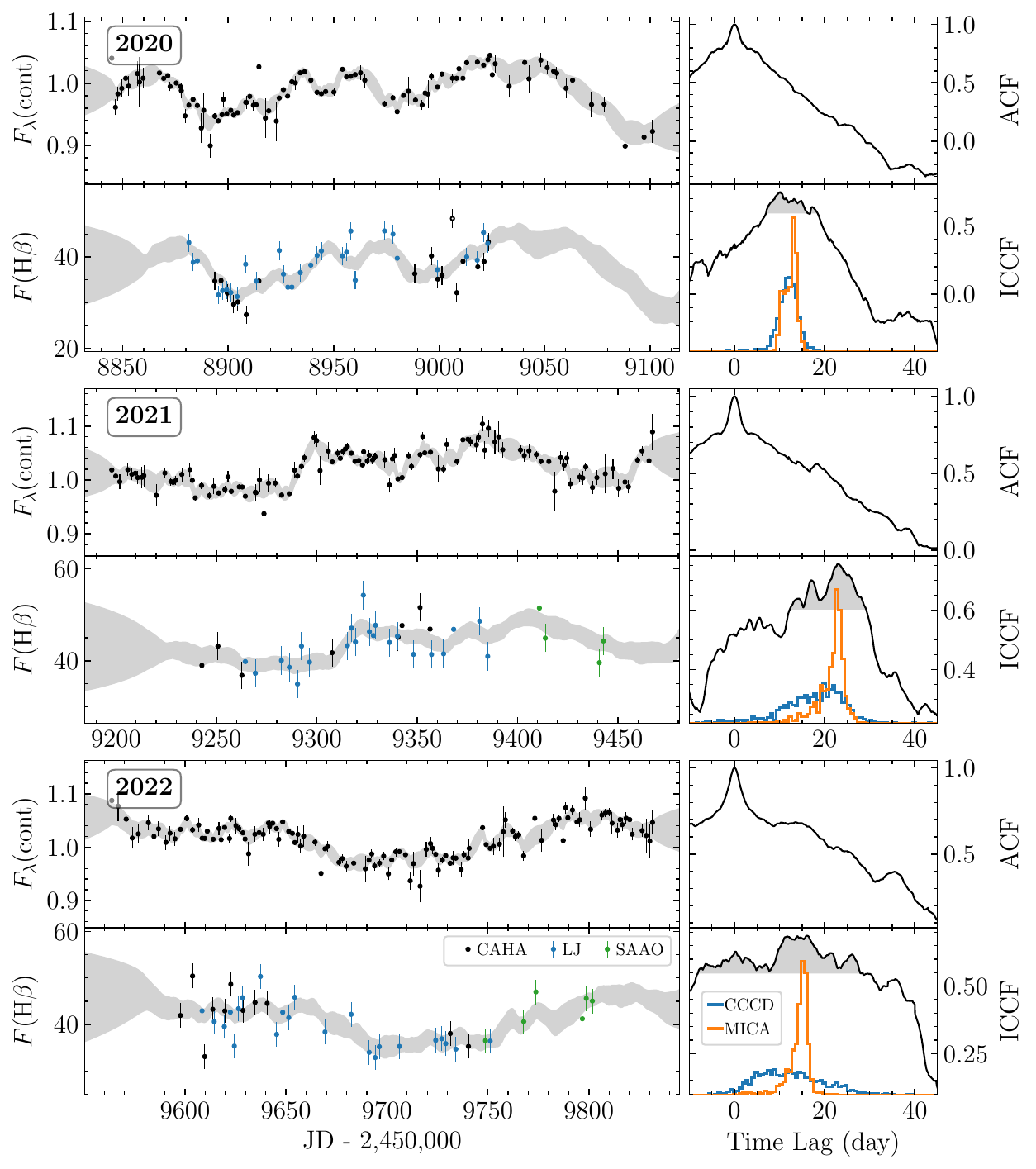}
\caption{Time lag measurements for IC 4329A. The left panels show the photometric (top)
and H$\beta$ (bottom) light curves for each observing season. The photometric light curves are converted from magnitudes, normalized
with their means and rebinned every one day. The gray shaded bands represent reconstruction using MICA.
In the panel for H$\beta$ light curves, the black, blue, and green points represent the observed data from CAHA, Lijiang, and SAAO, respectively. The H$\beta$ fluxes are in a unit of $10^{-13}~{\rm erg~s^{-1}~cm^{-2}}$.
The right panels show ACF  of the photometric light curves (top) and ICCFs between the photometric and H$\beta$ light curves (bottom).
The blue histograms show the cross-correlation centroid distributions (CCCDs) from ICCF and yellow histograms show the posterior distribution
of time lags from MICA. The grey shaded areas outline the region with the ICCF above 80\% of the peak value, which is used
for calculating the centroid time lag. In 2020 season, the H$\beta$ data point at JD~2,459,006 marked by an open circle is treated as an outlier (due to the telescope tracking issue) and excluded in time lag analysis.}
\label{fig_ccflag_ic4329a}
\end{figure*}

\begin{figure*}
\centering
\includegraphics[width=0.8\textwidth]{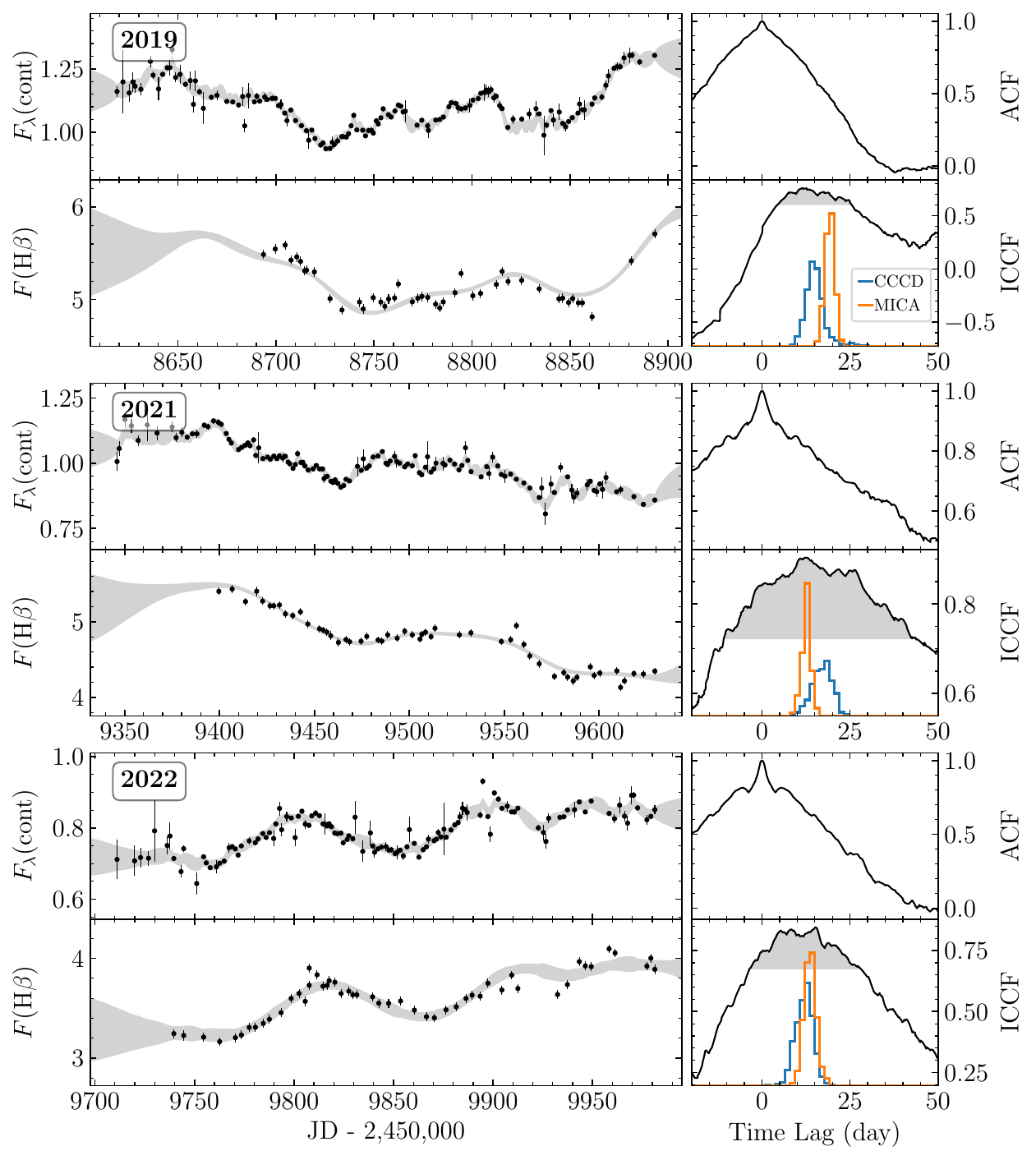}
\caption{Same as Figure~\ref{fig_ccflag_ic4329a}, but for Mrk 335. In the top left panel, the grey dashed line represents the linear trend used to detrend
the continuum light curve.}
\label{fig_ccflag_mrk335}
\end{figure*}

\begin{figure*}
\centering
\includegraphics[width=0.8\textwidth]{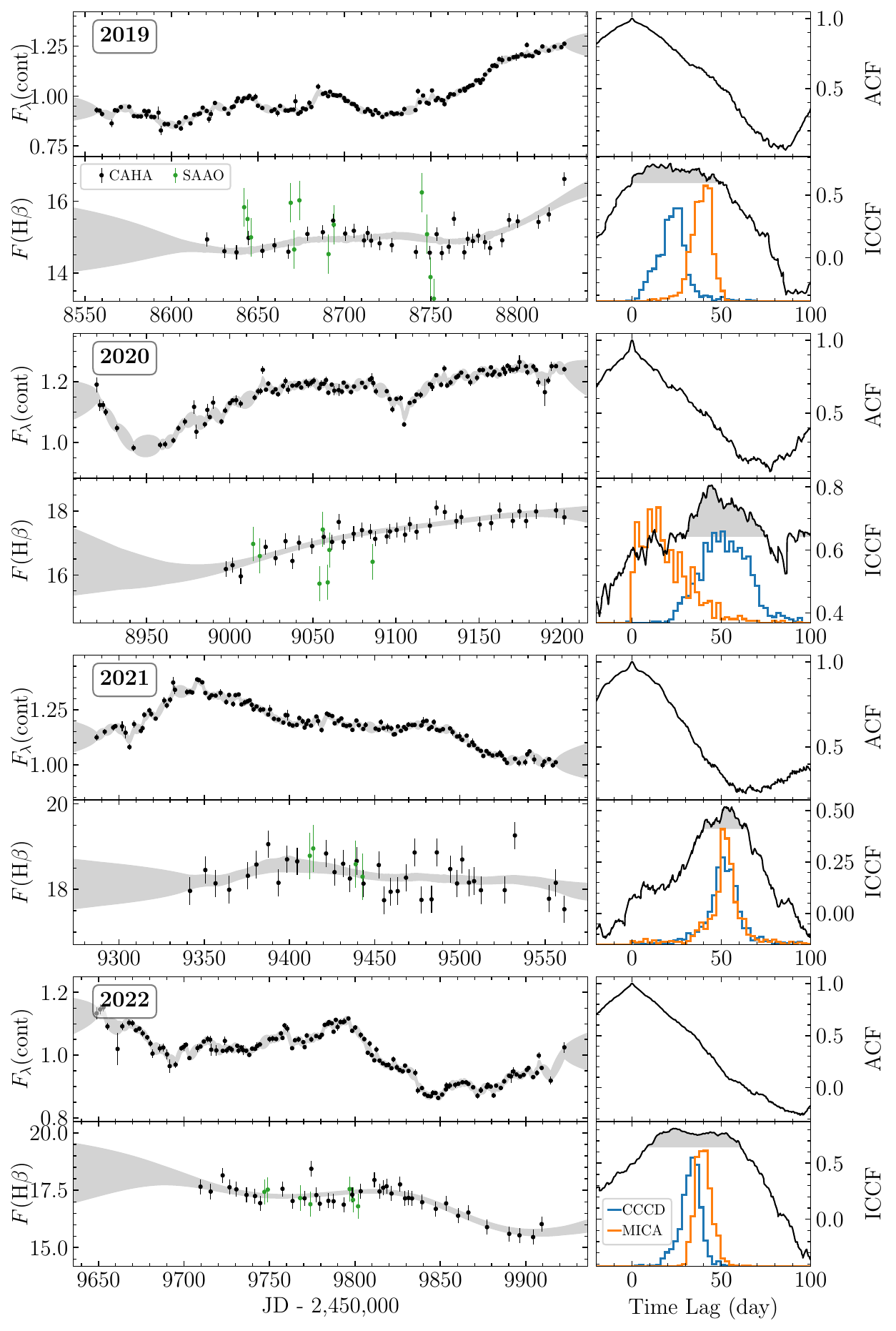}
\caption{Same as Figure~\ref{fig_ccflag_ic4329a}, but for Mrk 509. In season 2020, the H$\beta$ light curve has no
large variation structure and the obtained time lags are unreliable. Note that the SAAO data (green points) are not included in time lag analysis due to the
large scatter.}
\label{fig_ccflag_mrk509}
\end{figure*}

\begin{figure*}
\centering
\includegraphics[width=0.8\textwidth]{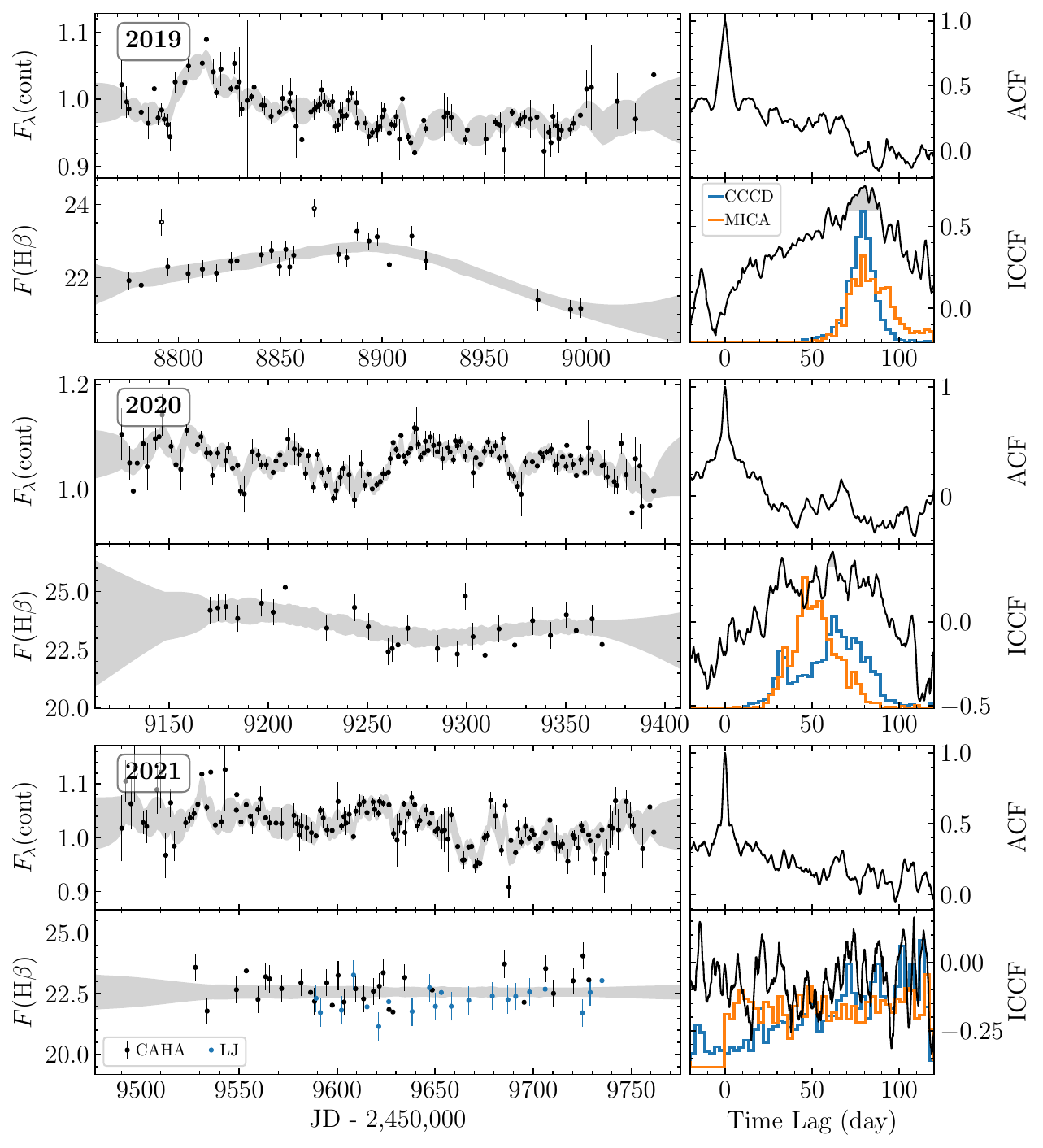}
\caption{Same as Figure~\ref{fig_ccflag_ic4329a}, but for Mrk 1239. There is not a reliable time lag in season 2021 due to small variability
in the H$\beta$ light curve. In 2019 season, the H$\beta$ data points at JD~2,458,792 and 2,458,866 marked by open circles are treated as outliers (due to poor weather conditions) and excluded in time lag analysis.}
\label{fig_ccflag_mrk1239}
\end{figure*}

\begin{figure*}
\centering
\includegraphics[width=0.8\textwidth]{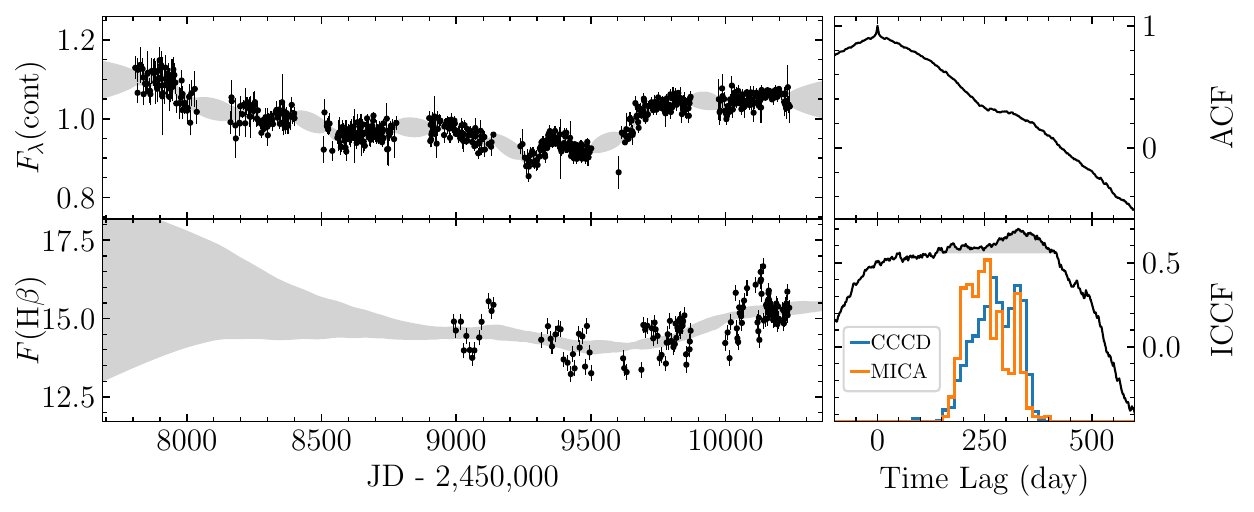}
\caption{Same as Figure~\ref{fig_ccflag_ic4329a}, but for PDS 456.}
\label{fig_ccflag_pds456}
\end{figure*}

In what follows we provide information on extra special treatments and comments for individual objects.

\paragraph{IC 4329A} This object displays a very red spectral shape. Through broad-band continuum modelling,
the previous study by \cite{Mehdipour2018} showed that IC~4329A has
an intrinsic reddening $A_V\approx2.0\pm0.2$ mag based on a flat extinction curve similar to that derived by \cite{Czerny2004}.
As a comparison, the foreground Galactic extinction is $A_V=0.159$ (\citealt{Schlafly2011}). We therefore additionally use the extinction law of
\cite{Czerny2004} to correct our observed spectra after applying the Galactic extinction correction.
Furthermore, IC 4329A shows a prominent spectral trough around 5150-5200 {\AA}, which is ascribed to
the Mg~$b$ triplet absorption ($\sim$5175~\AA) in the host galaxy's star light. We artificially magnify the fitting weights around this trough region
by trebling
the corresponding residuals. This allows us to better constrain the host galaxy component in spectral fitting.
The extra narrow window is set to (4190, 4280) {\AA}.

\paragraph{Mrk 335} This object has relatively narrow widths of the broad emission lines and shows weak extinction. It is impossible to reliably
separate the narrow H$\beta$ component and we use two Gaussians to model the whole H$\beta$ line. The extra narrow window is set to (4150, 4250) {\AA}.

\paragraph{Mrk 509} Unlike other objects, we use one Gauss-Hermite function plus one Gaussian to model the H$\beta$ line of Mrk 509.
There is a broad H$\beta$ wing extending to the [\ion{O}{3}] region. Using a Gauss-Hermite function helps to
alleviate the degeneracy between H$\beta$ and [\ion{O}{3}] parameters and make the spectral fitting more stable.
We use an extra window (4200, 4250) {\AA}, which is narrower than
that used for Mrk 335, in consideration of much broader line widths of Mrk 509. A narrower window helps to avoid the H$\gamma$ and H$\delta$ lines,
which are not included in our spectral fitting.
Moreover, Mrk 509 is bright and its spectra do not display a substantial host galaxy component.

\paragraph{Mrk 1239} For this object, an additional blue-shifted [\ion{O}{3}] $\lambda\lambda$4959, 5007 component with a typical velocity of
1200~km~s$^{-1}$ is required, which might arise from
outflows in the narrow-line region (e.g., \citealt{Pan2021}). Moreover, like Mrk 335, the H$\beta$ line is relatively narrow and it is not possible to reliably separate the narrow
component. We therefore use two Gaussians to model the whole H$\beta$ line and no longer separately add the narrow H$\beta$ component.
We use an extra narrow window of (4150, 4250) {\AA}, similar to that of Mrk 335.

\paragraph{PDS 456} This object does not show the [\ion{O}{3}] doublet, possibly indicating an absence of (or very weak) narrow H$\beta$ line.
We thus do not add the narrow H$\beta$ component. In addition, there appear absorption troughs around 4660, 4975, 5300~{\AA} and their origins are not
yet known. We use one Gaussian to model these absorptions and correct the trough around 4975 {\AA} when extracting the H$\beta$ line.
Similar to Mrk~509, there was no host galaxy component detected in our spectra of PDS 456. The extra narrow window is set to (4180, 4250) {\AA}.

\section{Light Curves and Time Lag Measurements}
\subsection{Photometric and Spectroscopic Light Curves}
In Section~\ref{sec_photometry}, we intercalibrate photometry from different datasets and obtain the combined
photometric light curves. The photometric cadences are overall much higher than spectroscopic monitoring cadences.
Therefore, in the following reverberation analysis, we use photometric light curves as the driving continuum ones.
To suppress noise, we further rebin the photometric data for every individual day.
Since reverberation analysis only depends on variations of light curves, rather than the absolute fluxes,
we convert the magnitudes into flux densities and simply normalize the resulting light curves with their means.

After spectral decomposition, we integrate the extracted H$\beta$ line profiles to obtain H$\beta$ line fluxes.
The integration window is adopted to center around 4861~{\AA} and vary among objects according to the line width.
The right panels of Figure~2 show the integration windows with vertical shaded bands.
In Figures~\ref{fig_ccflag_ic4329a}-\ref{fig_ccflag_pds456}, we plot the normalized continuum light curves and
the obtained H$\beta$ light curves. The H$\beta$ flux errors obtained from propagation of photon-counting errors
are usually quite small and do not account for any sources of systematic uncertainties such as slit positioning, poor weather
condition, etc. These systematics cause large abrupt flux fluctuations or scatter among adjacent epochs, which are difficult to interpret by intrinsic AGN variability. 
Following the procedure in \cite{Du2014}, we estimate systematic errors
by calculating the standard deviation of the residuals between the H$\beta$ flux data and the median-filtered light curve.
The filter window (in units of time) is set to twice the median sampling cadence of the H$\beta$ light curve.
This systematic error is finally added in quadrature to the original data errors. Table~\ref{tab_lc} lists the
estimated systematic errors for each season as well as for the whole monitoring period.

In Table~\ref{tab_lc}, we calculate the mean magnitude and fractional variability $F_{\rm var}$ of the photometric light curves of each object, as well as the mean flux and $F_{\rm var}$ of the 5100~{\AA} and H$\beta$ light curves.
Here, $F_{\rm var}$ is defined as (e.g., \citealt{Edelson2002})
\begin{equation}\label{eqn_fvar}
F_{\rm var} = \frac{1}{\langle F \rangle} \sqrt{\sigma_F^2 - \langle \sigma_{\rm err}^2\rangle},
\end{equation}
where $\langle F \rangle$, $\sigma_F^2$, and $\langle \sigma_{\rm err}^2\rangle$ are the mean flux,  flux variance, and mean square error, respectively.
Compared to the conventional definitions, we additionally include weights of data points to account for different measurement errors, namely,
\begin{gather}
\langle F \rangle = \sum_i w_i F_i,\\
\sigma_F^2 = \frac{1}{1-\sum_i w_i^2} \sum_i w_i\left(F_i - \langle F \rangle\right)^2,\label{eq_var}\\
\langle \sigma_{\rm err}^2\rangle = \sum_i w_i \sigma_i^2,
\end{gather}
where the weight $w_i$ is assigned as the inverse square error of each point, namely,
\begin{equation}
 w_i = \frac{1/\sigma_i^2}{\sum_i(1/\sigma_i^2)},
\end{equation}
where $\sigma_i$ is the error of $i$-th point. The factor before the summation in the right-hand side of Equation~(\ref{eq_var})
comes from a bias correction to the regular weighted variance.

\begin{figure}
\centering
\includegraphics[width=0.48\textwidth]{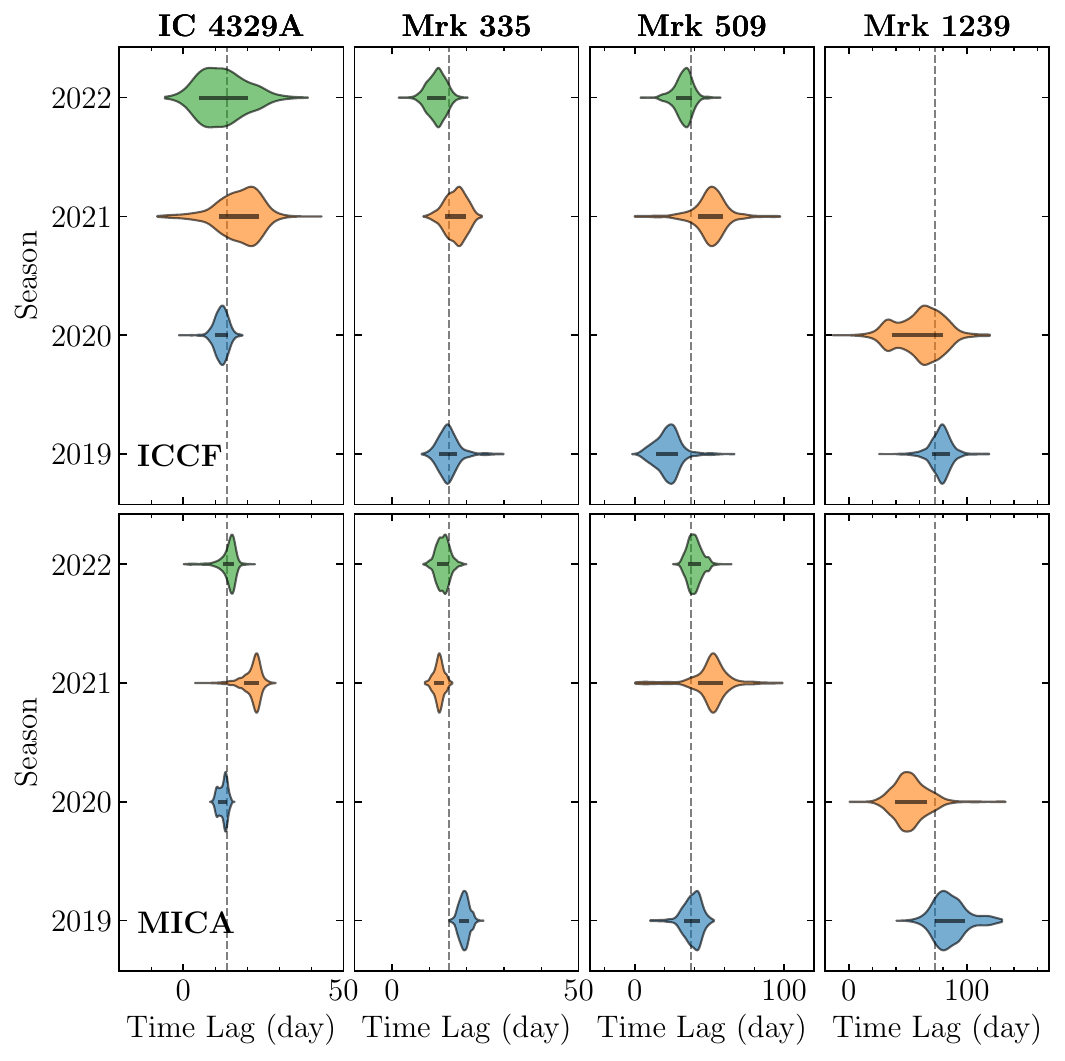}
\caption{A comparison of H$\beta$ time lags between seasons from the ICCF (top) and MICA (bottom) methods.
The horizontal black lines inside the violin plots represent the 68.3\% confidence intervals. Vertical dashed lines
represent the means of all ICCF and MICA lags and are plotted to guide the eye.}
\label{fig_lag_compare}
\end{figure}

\subsection{Time Lag Measurements}\label{sec_lag}
We calculate time lags between the continuum and H$\beta$ line fluxes using two methods. One is the standard
interpolated cross-correlation function (ICCF) method (\citealt{Gaskell1986, Gaskell1987}) and the other is the {\textsc{MICA}}\footnote{The living code for MICA2 is publicly available at \url{https://
github.com/LiyrAstroph/MICA2}, while the version used in this work is available at \url{https://doi.org/10.5281/zenodo.11082109}.} method (\citealt{Li2014}).
As usual, for the ICCF method, the time lag is assigned as the centroid of the ICCF above 80\% of the peak value. The associated
uncertainties are estimated by 68.3\% quantiles of the cross-correlation centroid distribution, generated from Monte Carlo simulations with
the ﬂux randomization and random subset selection (FR/RSS) method (\citealt{Peterson1998b}). The {\textsc{MICA}} method uses the damped random walk
process to describe the light curve variability and a series of displaced Gaussian family to model the transfer function. {\textsc{MICA}}
employs the diffusive nested sampling algorithm (\citealt{Brewer2011}) to explore the posterior probability distribution and
generate Markov chains of parameters. The algorithm belongs to a variant of the nested sampling algorithm,
and is therefore able to directly calculate the Bayesian evidence (e.g., see \citealt{Sivia2006}).
The best number of Gaussians can be determined by comparing the Bayesian evidence among different numbers of Gaussians.
Here, for the sake of comparison with ICCF results, we only use one Gaussian and determine the time lag as the median Gaussian center and quote its
uncertainties from 68.3\% quantiles of the posterior sample.

Figures~\ref{fig_ccflag_ic4329a}-\ref{fig_ccflag_pds456} show auto-correlation functions (ACFs) and ICCFs of the light curves as well as time lag distributions from ICCF
and MICA methods for each season of each object.Table~\ref{tab_lag} summarizes these analysis results.  Figure~\ref{fig_lag_compare} compares the H$\beta$ time lags obtained over the monitoring seasons using violin plots. As can be seen, the ICCF and MICA time lags are generally consistent with
each other (within 1-2$\sigma$ uncertainties), except that MICA time lags have relatively smaller uncertainties. This is because the MICA method presumes the form of transfer function and is less sensitive to sampling gaps. Besides, as demonstrated by \cite{Peterson1998b}, the FR/RSS procedure in the ICCF method usually yields conservative error estimates and the quoted errors might be larger than real uncertainties (see also \citealt{Yu2020}).
The maximum cross-correlation coefficient $r_{\rm max}$ is overall larger than 0.5, except for Mrk 1239,
which has the lowest variability in both continuum and H$\beta$ line among the five objects ($F_{\rm var}\lesssim3\%$).

During running MICA, we identified one outlier point (at JD~2,459,006) in the H$\beta$ light curve of IC 4329A, which causes an abnormally small time lag uncertainty. In the 2019 season of Mrk 1239, we also identified two outlier points (at JD~2,458,792 and 2,458,866) in the H$\beta$ light curve, which lead to an abnormal time lag distribution. These points are marked by open circles in Figures~\ref{fig_ccflag_ic4329a} and \ref{fig_ccflag_mrk1239}, respectively. We checked the raw data and observational logs and confirmed that the corresponding exposures were subject to either the instrumental issues or poor weather conditions. We therefore excluded them in time lag analysis. In addition, we are unable to find a time lag in the 2022 season of Mrk 1239 because of lacking significant variation patterns in the H$\beta$ light curve,
which is overwhelmed by noises and seems to stay at a nearly constant level during the whole season.
In the 2020 season of Mrk~509, the H$\beta$ flux keeps continuously rising without showing a variation structure, so the obtained
time lags are not reliable for both ICCF and MICA and thus not listed in Table~\ref{tab_lag}.

In Table~\ref{tab_lag}, we also list the standard deviation of the Gaussian transfer function obtained from MICA. This value approximately reflects
spatial extension of the H$\beta$ BLR. IC~4329A has the narrowest BLR extension (relative to the BLR size $R=c\tau$), whereas other objects
have the BLR extension comparable to the BLR size, namely, $w_{\rm mica}\approx\tau_{\rm mica}$. It is not known why IC 4329A is different from other
objects, but comparing the H$\beta$ profile shows that only IC 4329A displays a prominent double-peaked H$\beta$ profile.

In the above analysis, we measure time lags for individual seasons separately for the four objects (IC 4329A, Mrk 335, Mrk 509, and Mrk 1239). We do not perform analysis on the full-span light curves considering the possibility that the BLR kinematics might change in the dynamical timescale, which can be estimated as
\begin{eqnarray}\label{eqn_tdyn}
t_{\rm dyn} &=& \sqrt{\frac{R_{\rm BLR}^3}{GM_\bullet}}=f_{\rm BLR}^{-1/2}
\frac{R_{\rm BLR}}{\Delta V} \nonumber\\
&\approx&4\,{\rm yr}\,f_{\rm BLR}^{-1/2}\left(\frac{R_{\rm BLR}}{10\,{\rm light\text{-}day}}\right)\left(\frac{\Delta V}{2000\,{\rm km/s}}\right)^{-1},
\end{eqnarray}
where $\Delta V$ is width of the emission line. The resulting dynamical timescales for the four objects are on the order of years, meaning that time lag analysis over a long period might be affected by changes in the BLR kinematics. However, for PDS 456, the dynamical timescale is roughly 65 yr, much longer than our monitoring period. Therefore, it is reasonable to conduct time lag analysis using the full-span light curves for PDS 456.
For the sake of completeness, we present the MICA results with all seasons for the other four objects in Appendix~\ref{app_whole}.

Below we comment on time lag measurements of each object individually.

\paragraph{IC\,4329A} When comparing seasons, the continuum flux of IC 4329A tends to stay stable and does not have large seasonal
changes. The H$\beta$ light curves are subject to mild scatter and the time lags from ICCF bear relatively large uncertainties.
The time lags for different seasons are consistent within 1$\sigma$ uncertainties from the ICCF and within 2$\sigma$ uncertainties from the MICA method.

\paragraph{Mrk\,335} It shows the largest seasonal changes in continuum and H$\beta$ fluxes, which gradually decrease from 2019 to 2022 seasons
by $\sim$25\%. In the 2019 season, there appears a visible difference in the long-term trends
between the continuum and H$\beta$ light curves (see the top panels in Figure~\ref{fig_ccflag_mrk335}). The continuum flux slowly increases with time
whereas the H$\beta$ line tends to stay at a constant flux level. This leads to a difference (at 1.5$\sigma$ confidence level) in the time lags measured by ICCF and MICA.
It is worth mentioning that the results come to an agreement if we follow the conventional manipulation to detrend the light curves using a linear polynomial (e.g., see \citealt{Denney2010, Peterson2014, Li2020}). However, considering that the physical explanation for detrending manipulation is still elusive, we therefore use the measurements without
detrending and defer the detailed investigation of the apparently different trends to a separate work (Chen et al. 2024, in preparation).
In the 2021 season, the ICCF and MICA time lags are consistent within 1$\sigma$ uncertainties, although the peaks of the two distributions are slightly different
(see the middle panels in Figure~\ref{fig_ccflag_mrk335}).

\paragraph{Mrk\,509} In the 2019 season of Mrk 509, the H$\beta$ time lag from MICA is marginally larger than that from ICCF at 1.8$\sigma$ confidence level (see the top panels of Figure~\ref{fig_ccflag_mrk509}). We ascribe this difference to the different variation trends in the continuum and H$\beta$ light curves. During the period JD~2,458,600-2,458,800, the continuum flux gradually rises by a factor of $\sim30$\%, whereas the H$\beta$ flux almost maintains a constant level.
The ICCF is mainly contributed by the minor bump around JD~2,458,700 and the rapidly rising feature after JD~2,458,800 in the H$\beta$ light curve.
Instead, MICA tends to reproduce all variation features under the condition that the H$\beta$ variations are blurred and delayed echoes to the continuum variations.
The broad minor trough around JD~2,458,750 in the H$\beta$ light curve is not well fitted. Nevertheless, considering the obtained time lags in the other seasons,
a H$\beta$ time lag of $\sim40$ days from MICA might be more plausible in the 2019 season.

\paragraph{Mrk\,1239} It shows the lowest variability in both continuum and H$\beta$ light curves. This leads to quite large uncertainties of the obtained
time lags for both ICCF and MICA. Despite the overall amplitudes of continuum variability being comparable in the three seasons, we note that, unlike the other two seasons,
the 2021 season does not exhibit a significant variation pattern in the first half period. As a result, the large measurement errors overwhelm the H$\beta$ variations
and make us unable to detect the time lag.

\paragraph{PDS\,456} It shows some fine variation structures in the H$\beta$ light curve that seem not to have counterpart structures in
the continuum light curve (see Figure~\ref{fig_ccflag_pds456}). Both ICCF and MICA methods find the time lag between the long-term variation patterns of
the H$\beta$ and continuum light curves. The obtained time lag of $\sim$250-280 days together with a broad Gaussian standard deviation of $w_{\rm mica}\sim$120 days means that any variations with timescale shorter than $w_{\rm mica}$ in the H$\beta$ flux will be blurred out. It is not clear why PDS 456 exhibits non-echoing H$\beta$ variations.

\begin{deluxetable}{cccc}
\tabletypesize{\footnotesize}
\tablecaption{Rest-frame Wavelength Windows Used for Measuring H$\beta$ fluxes and Subtracting Continuum Residuals in RMS Spectra. \label{tab_window}}
\tablehead{\colhead{~~~~~~~~~Name~~~~~~~~~}  & \colhead{~~~~~~~~~H$\beta$~~~~~~~~~}  &  \multicolumn{2}{c}{Continuum}\\\cline{3-4}
                           &                     &  \colhead{~~~~~~~~~Left~~~~~~~~~}   & \colhead{~~~~~~~~~Right~~~~~~~~~}\\
                           &  \colhead{(\AA)}      &  \colhead{(\AA)}          & \colhead{(\AA)}
}
\startdata
IC 4329A & 4750-4950   & 4700-4730 & 5085-5115\\
Mrk 335   & 4800-4920  & 4770-4800 & 5085-5115 \\
Mrk 509   & 4760-4940  & 4700-4730 & 5085-5115\\
Mrk 1239  & 4800-4920  & 4770-4800 & 5135-5165\\
PDS 456   & 4740-4980  & 4600-4630 & 5085-5115
\enddata
\end{deluxetable}

\begin{figure*}[ht!]
\centering
\includegraphics[height=0.45\textwidth]{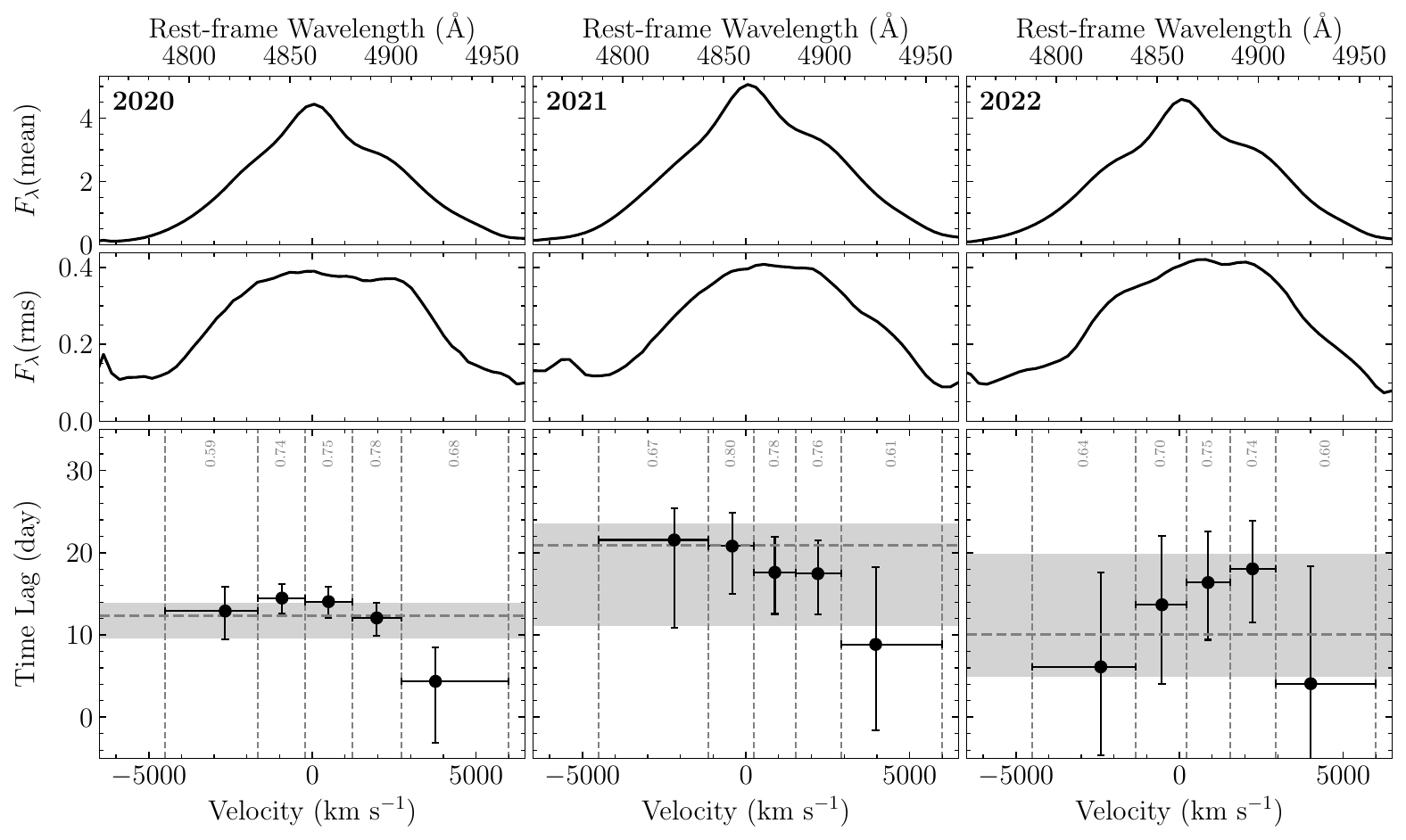}
\caption{(Top) mean spectra, (middle) rms spectra,  and (bottom) velocity-binned time lags (in the observed frame) for IC 4329A in seasons 2020-2022.
In the bottom panels, the numbers in grey represent the peak ICCF coefficient $r_{\rm max}$ in each velocity bin.
In each bin, the point's velocity is assigned as the centroid velocity weighted by the rms fluxes.
The dashed horizontal lines with grey shaded bands show the time lags and uncertainties from the integrated H$\beta$ light curves.
The grey open points show velocity bins with $r_{\rm max}<0.5$, which indicates that the time lags might be unreliable.
The flux densities are in units of  $10^{-14}~\rm erg~s^{-1}~cm^{-2}~\text{\AA}^{-1}$.}
\label{fig_vel_res_lag_ic4329a}
\end{figure*}

\begin{figure*}[ht!]
\centering
\includegraphics[height=0.45\textwidth]{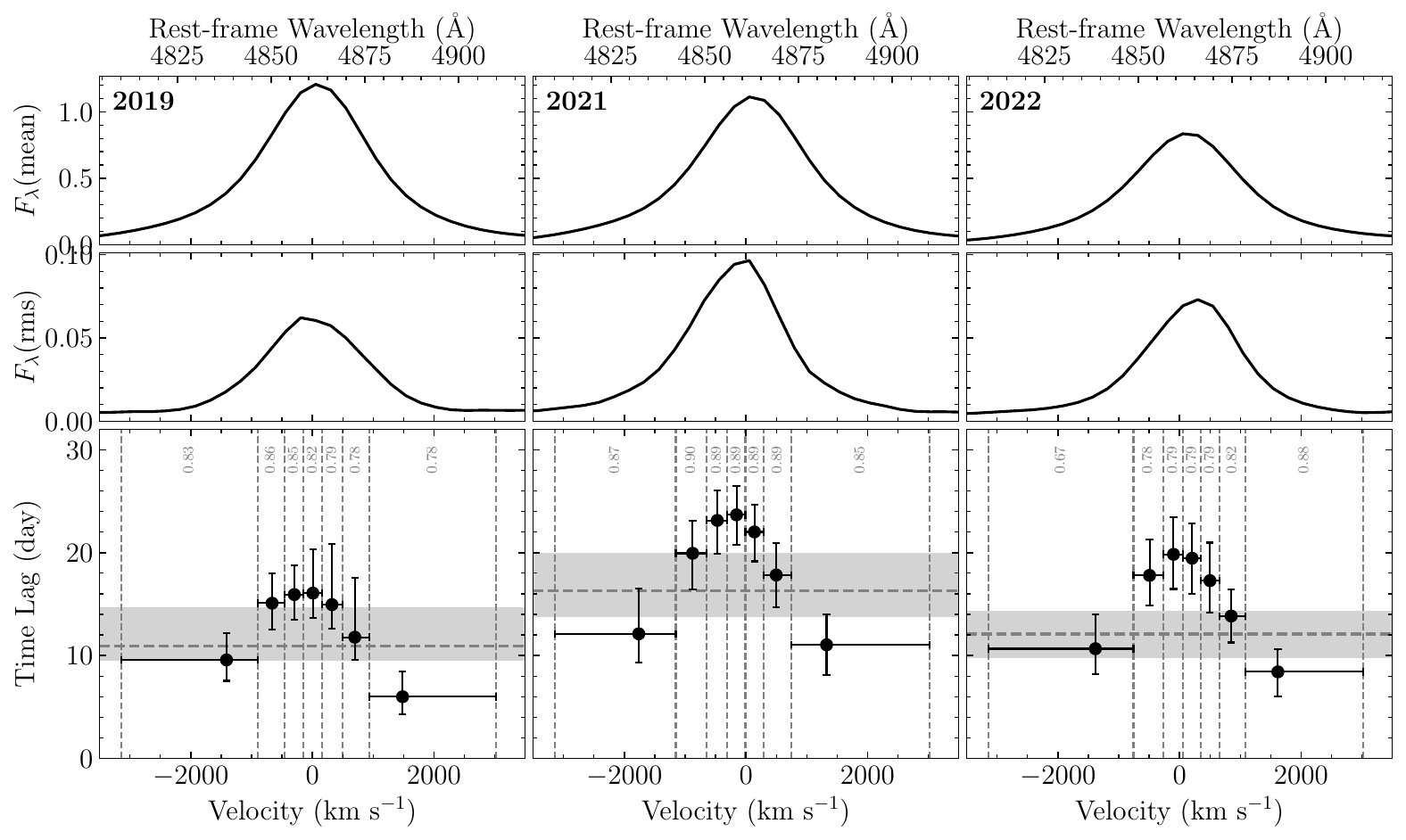}
\caption{Same as Figure~\ref{fig_vel_res_lag_ic4329a} but for Mrk 335 in seasons 2019, 2021, and 2022.}
\label{fig_vel_res_lag_mrk335}
\end{figure*}

\begin{figure*}
\centering
\includegraphics[height=0.45\textwidth]{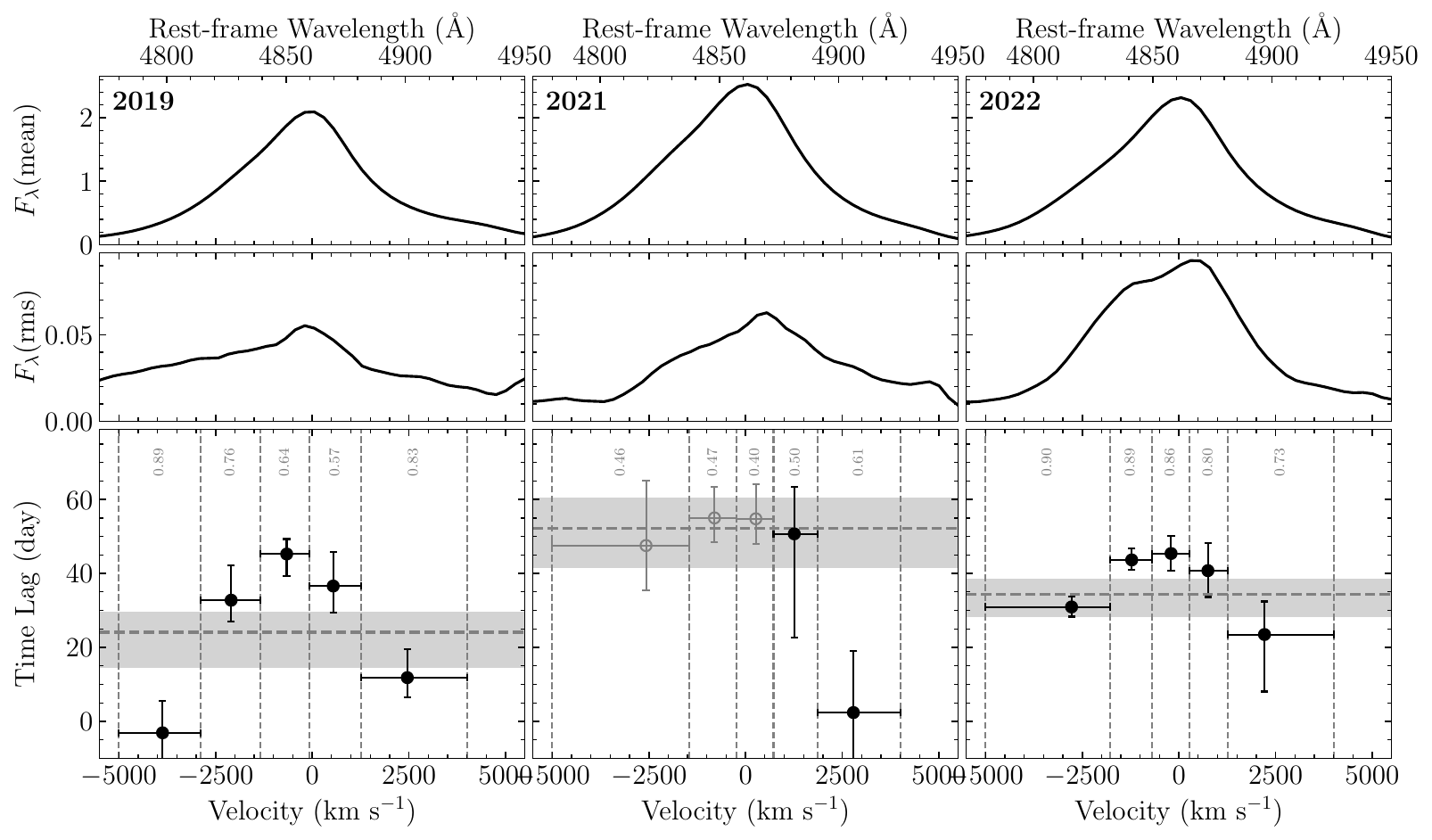}
\caption{Same as Figure~\ref{fig_vel_res_lag_ic4329a} but for Mrk 509 in seasons 2019, 2021, and 2022.}
\label{fig_vel_res_lag_mrk509}
\end{figure*}

\begin{figure*}[t!]
\centering
\includegraphics[height=0.45\textwidth]{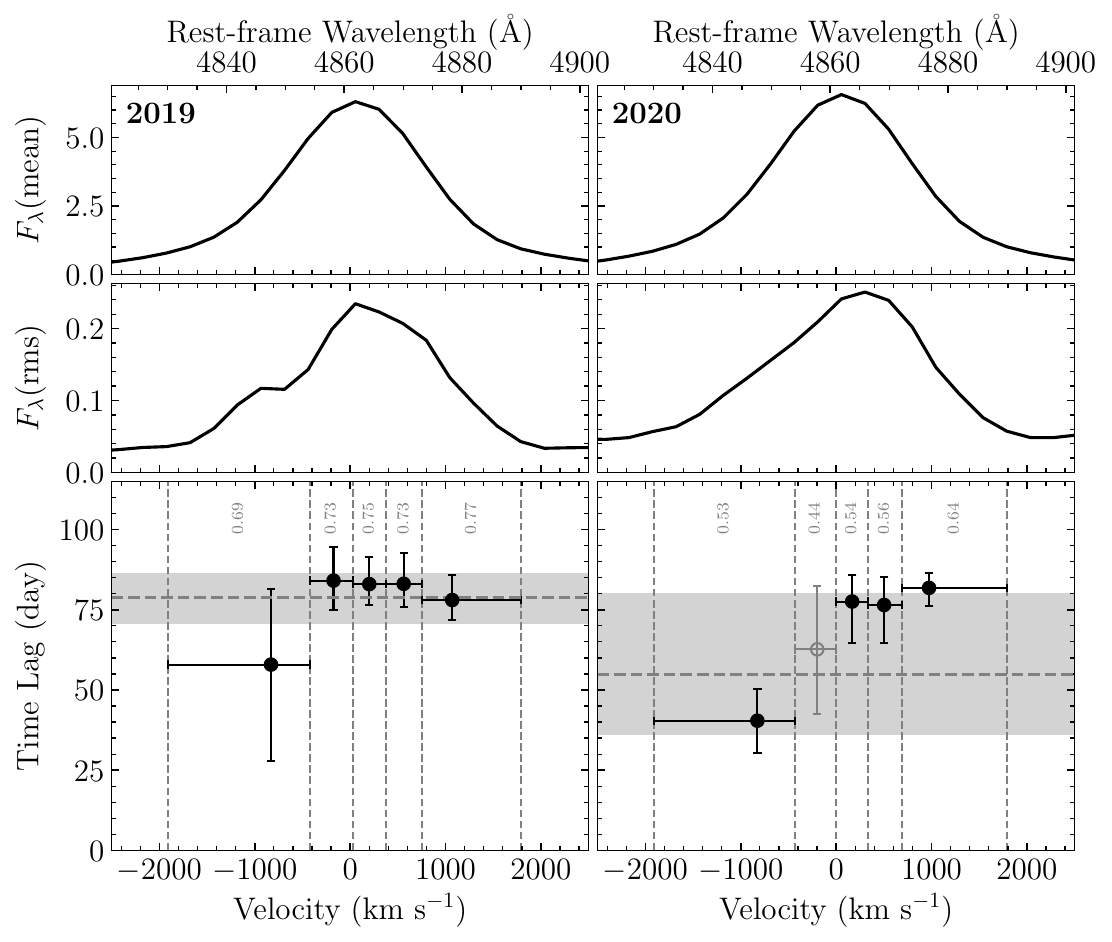}
\caption{Same as Figure~\ref{fig_vel_res_lag_ic4329a} but for Mrk 1239 in seasons 2019 and 2020.}
\label{fig_vel_res_lag_mrk1239}
\end{figure*}

\subsection{Velocity-resolved Time Lag Measurements}
\begin{figure}[t!]
\centering
\includegraphics[height=0.45\textwidth]{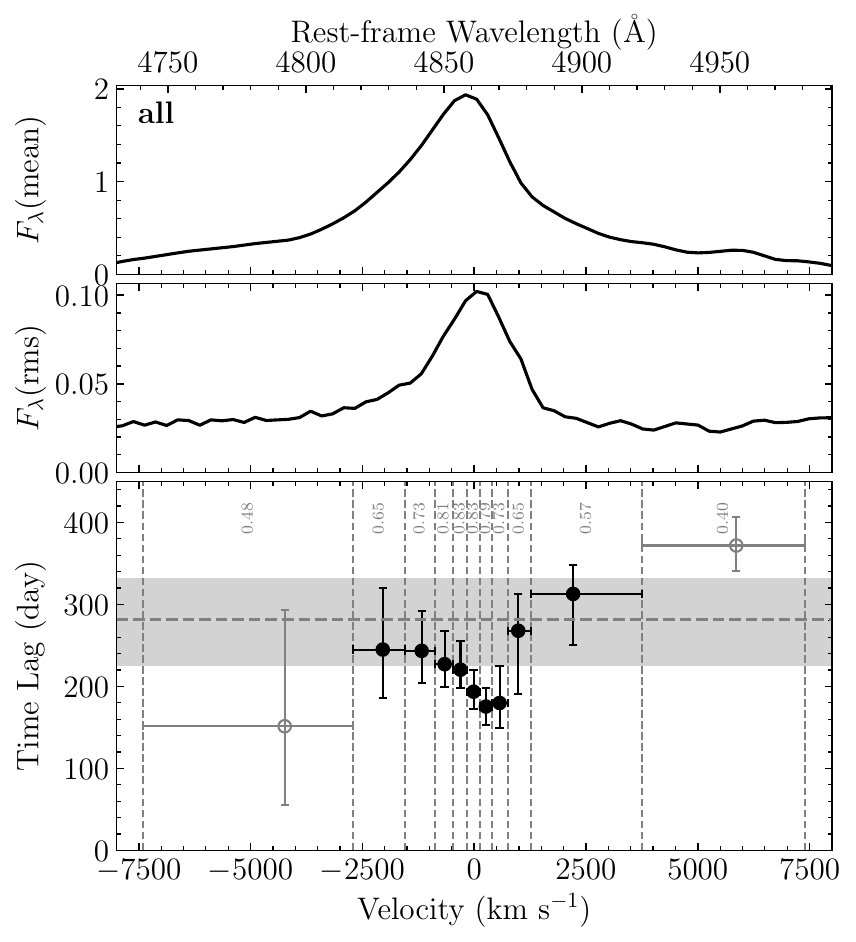}
\caption{Same as Figure~\ref{fig_vel_res_lag_ic4329a} but for PDS 456 over the whole monitoring period.}
\label{fig_vel_res_lag_pds456}
\end{figure}

The quality of the present dataset allows us to measure velocity-resolved time lags, which can provide information about the BLR kinematic features.
For each object, we divide the H$\beta$ rms spectrum in each season into 5-11 velocity bins and ensure that all bins have approximately equal rms fluxes.
The adopted wavelength windows are listed in Table~\ref{tab_window} and are kept the same for all seasons.
We integrate the fluxes in each velocity bin and obtain the corresponding light curve. We then employ the same ICCF method as described in
Section~\ref{sec_lag} to measure the time lags and uncertainties with respect to the continuum light curve.
Figures~\ref{fig_vel_res_lag_ic4329a}-\ref{fig_vel_res_lag_pds456} show the obtained velocity-resolved time lags for the five objects, along with
their mean and rms spectra in each season.
Again, we do not detect reliable time lags in the 2020 season of Mrk~509 because of the lack of large variation patterns and in
the 2021 season of Mrk 1239 because of the low H$\beta$ variability. These two seasons are therefore not included in the figures.

The velocity-resolved time lags of IC 4329A in Figure~\ref{fig_vel_res_lag_ic4329a}
display seemingly diverse features among seasons 2020, 2021, and 2022. However, because of the large time lag uncertainties, we cannot rule out the possibility that the velocity-resolved time lags in the three seasons might be consistent with each other. Moreover, we also cannot distinguish between disk-like BLRs, inflows/outflows, and thick spherical BLRs. The latter case procedures a flat distribution of time lags with velocity around the line core.
IC~4329A was recently monitored by \cite{Bentz2023} between March 7 and July 31, 2022, partially overlapping with our monitoring season 2022.
They reported a H$\beta$ centroid time lag of $16.33_{-2.28}^{+2.59}$ days, consistent within uncertainties with our measurements.
Their measured velocity-resolved time lags imply a disk-like signature, that is, the time lag peaks in line core and decline towards the both
red and blue line wings. This might align with the double-peaked H$\beta$ line profile in IC~4329A, which is commonly explained by a disk-like BLR
structure. Nevertheless, we stress that both our and \cite{Bentz2023}'s results bear large uncertainties in the obtained time lags, therefore, more sophisticated analysis and/or additional RM campaigns are worthwhile to reliably determine the structure and kinematics of the BLR.

Both Mrk 335 (Figure~\ref{fig_vel_res_lag_mrk335}) and Mrk 509 (Figure~\ref{fig_vel_res_lag_mrk509}) clearly show a disk-like signature over the three seasons (2019, 2021, and
2022). Previous studies of Mrk 335 by \cite{Grier2013} and \cite{Du2016} reported an asymmetric structure in velocity-resolved time lags, with longer lags
toward the blue wing and shorter lags toward the red wing. Such a feature is usually interpreted by inflowing gas in the BLR. However, we
do not find this feature in our three-season monitoring, indicating that the BLR kinematics in Mrk 335 might change from inflows into Keplerian disk-like motion.
Such kinematic changes are plausible considering that the typical BLR dynamical timescale is on the order of several years (see Equation~\ref{eqn_tdyn}).
We note that we used decomposed H$\beta$ profiles to calculate velocity-resolved lags, an approach different from that in \cite{Grier2013} and \cite{Du2016}, which used a straight line to subtract the underlying continuum and other components. In the latter approach, there will remain some residuals from other emission line components around the H$\beta$ line, such as \ion{He}{2}$\lambda$4686 and \ion{Fe}{2}, which might affect the H$\beta$ time lag measurements, especially in the wings. From the spectral decomposition shown in Figure~\ref{fig_specfit}, we can find that the contamination of
\ion{He}{2} and \ion{Fe}{2} lines to the H$\beta$ line is minimal. We thereby expect that the two approaches do not make significant difference in determining the
velocity-resolved H$\beta$ time lags.

Mrk 1239 (Figure~\ref{fig_vel_res_lag_mrk1239}) has relatively narrow line widths, slightly larger than the instrumental broadening,
resulting in similar lags around the H$\beta$ line core in the left panels of Figure~\ref{fig_vel_res_lag_mrk1239}.
In addition, the maximum ICCF coefficients $r_{\rm max}$ are below 0.7 across almost all velocity bins in seasons 2019 and 2020. This is because of
the low variability in both the continuum and H$\beta$ light curves, as shown in Figure~\ref{fig_ccflag_mrk1239}. As for PDS 456
(Figure~\ref{fig_vel_res_lag_pds456}), it seems that its rms spectra consists of a narrow core plus broad
wings (see also Figure~\ref{fig_mean_rms}).
The profile of the velocity-resolved lags is irregular, seeming hard to be explained by solely a disk or outflows.
In a nutshell, Mrk~1239 and PDS 456 might need further continuous monitoring with higher-resolution monitoring to confirm these results.

\begin{figure*}
\centering
\includegraphics[width=0.9\textwidth]{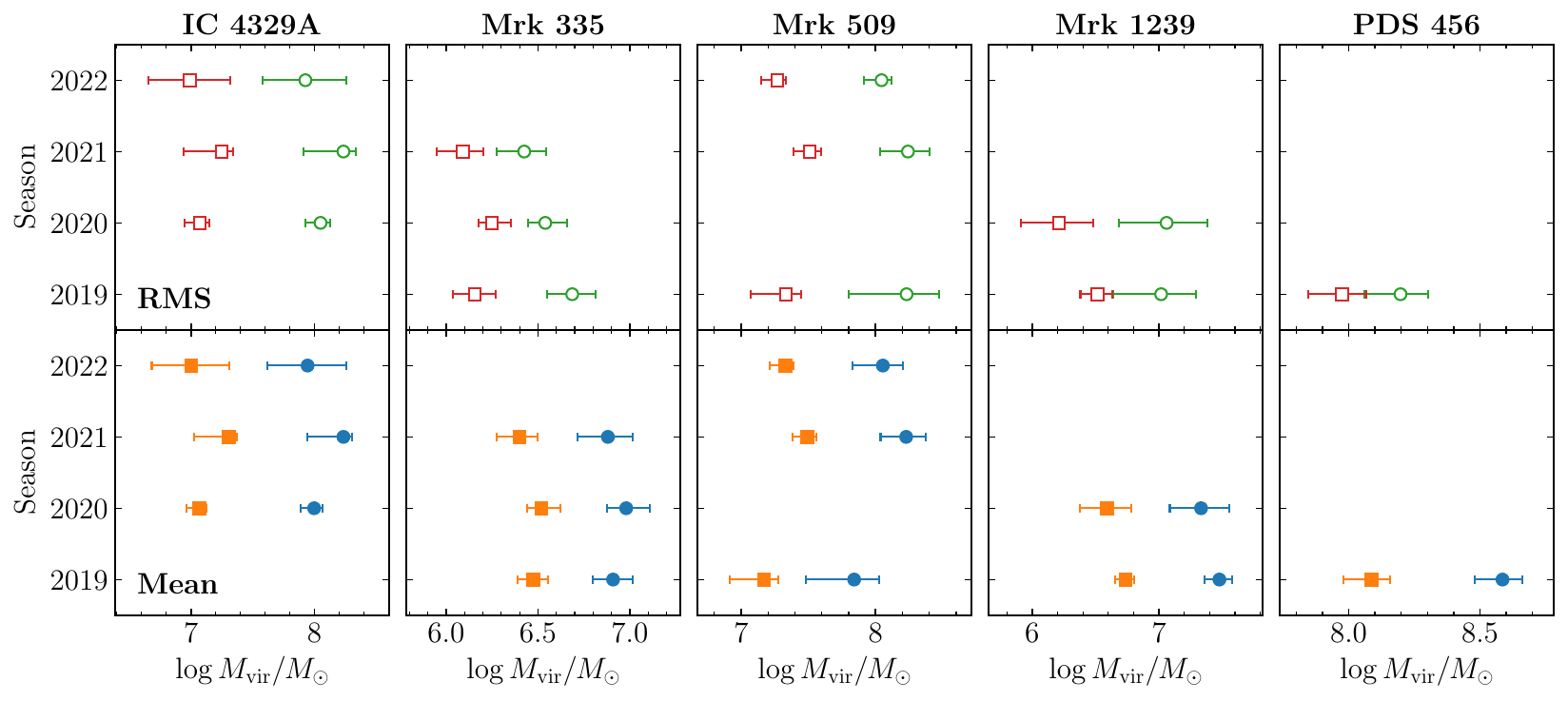}
\caption{The virial products ($M_{\rm vir}$) for different types of H$\beta$ line widths. Solid and open points are for mean and rms spectra, and circles and squares are for FWHM and line velocity dispersion, respectively. Note that the black hole mass can be obtained by multiplying $M_{\rm vir}$ with
a virial factor $f_{\rm BLR}$ (see Equation~\ref{eqn_mass}), which depends on the adopted measure of line width (e.g., \citealt{Ho2014}).}
\label{fig_mass}
\end{figure*}

\section{Broad H$\beta$ Line Widths and Black Hole Masses}
\subsection{Broad H$\beta$ Line Widths}
We measure broad H$\beta$ line widths, namely, full width at half maximum (FWHM) and line velocity dispersion from
the mean and rms spectra as follows. For each season, we calculate the mean and rms spectra of the extracted H$\beta$ profiles using Equations
(\ref{eqn_mean}) and (\ref{eqn_rms}) based on the spectral decomposition described in Section~\ref{sec_decomp}. Considering that
it is difficult to reliably decompose the narrow H$\beta$ component in most of our objects (except for IC 4329A, which has a distinct narrow bump in
the line core, and PDS 456, which does not have a significant narrow component), we keep both
the broad and narrow components when extracting the H$\beta$ profiles. This leads to the mean spectrum contaminated by the narrow component.
However, the rms spectrum is less sensitive to such a contamination because the narrow component is largely eliminated in calculating the variance
in Equation (\ref{eqn_rms}). Following \cite{Hu2015}, we take into account of the influence of the narrow component in line widths by artificially
assuming a flux ratio 10\% of the narrow H$\beta$ to [\ion{O}{3}]~$\lambda$5007 line. Specifically, we first subtract the narrow H$\beta$ component
from the extracted H$\beta$ profile in each epoch with the assumed narrow-line flux ratio before calculating the mean and rms spectra. The FWHM widths and line dispersions
are then obtained in a conventional way.

The uncertainties of line widths are assessed using the following procedure. First we consider uncertainties from
narrow H$\beta$ component subtraction. We calculate line widths in two cases: one is without narrow-line subtracting and the other is with subtracting a narrow H$\beta$ assumed to be 20\% of the [\ion{O}{3}]~$\lambda$5007 line flux. We assign the uncertainties as the half of the line width differences between these two extremes. The second source of uncertainties comes from the consideration that our finite number of observations
are just one possible realization of the true process and the measured width might be biased by applying only one realization.
This sort of uncertainties is estimated using the bootstrapping method (e.g., \citealt{Peterson1998a, Barth2015}).
Given a dataset with $N$ spectra, we construct a new realization by randomly selecting $N$ spectra with replacement, combine duplicated spectra (with the errors divided by root of the number of duplication), and then measure line widths from new mean and rms spectra.
Repeating this procedure 1000 times results in
a distribution of line widths, from which we assign the uncertainties as the standard deviation. We finally sum up the above two sources
of uncertainties in quadrature.

As a final step, we need to correct for the instrumental broadening to obtain the intrinsic line widths. The instrumental broadening
is estimated by comparing the line width of [\ion{O}{3}]~$\lambda$5007 of our spectra with that from high-resolution spectra observed by
\cite{Whittle1992}. For PDS 456 without showing the [\ion{O}{3}] doublet, we use the typical broadening width of $\sim1000$~km~s$^{-1}$ in FWHM (e.g., \citealt{Hu2021}).

The measured line widths are summarized in Table~\ref{tab_mass}. Regarding line widths of mean spectra, Mrk~509 and Mrk 1239 bear
large uncertainties arising from subtraction of the strong narrow line component. Conversely, PDS 456 does not show a narrow line component and therefore bears
the smallest uncertainties, only resulting from the bootstrapping. As mentioned above, line width uncertainties in rms spectra are
insensitive to narrow line subtraction, but instead mainly depend on the line variability amplitude as well as data quality.

\subsection{Black Hole Masses}
As usual, we estimate the black hole mass using the virial product as
\begin{equation}
M_\bullet = f_{\rm BLR} M_{\rm vir} =  f_{\rm BLR} \frac{c\tau\times (\Delta V)^2 }{G},
\label{eqn_mass}
\end{equation}
where $c$ is the speed of light, $G$ is the gravitational constant, $f_{\rm BLR}$ is the virial factor, $\tau$ is the time lag of the
emission line, and $\Delta V$ is the line width. From emission line profiles, it is clear that the BLR must have a certain spatial extension and
cannot be simply treated as point-like. The defined virial factor thereby incorporates all uncertainties and/or our ignorance on the detailed
structure and kinematics of BLRs.
The conventional approach to  statistically calibrate the virial factor is through comparing RM AGNs with measured stellar velocity dispersions against
the $M_\bullet-\sigma_\star$ relation of local inactive galaxies (e.g., \citealt{Onken2004, Ho2014, Yu2019}).

Table~\ref{tab_mass} lists the virial products for each type of line widths. Figure~\ref{fig_mass} compares the virial products among different line width types and different seasons. We find generally good consistency in virial products. For Mrk 335, it appears that rms spectra-based virial products are systematically smaller than mean spectra-based ones, this is because of the narrower line widths from rms spectra (see Figure~\ref{fig_vel_res_lag_mrk335}).
By using the FWHM widths from the mean spectra and a unity virial factor $f_{\rm BLR}=1$,
we obtain the black hole mass averaged over the seasons
$\log M_\bullet/M_\odot=8.02_{-0.14}^{+0.09}$, $6.92_{-0.12}^{+0.12}$, $8.01_{-0.25}^{+0.16}$, $7.44_{-0.14}^{+0.13}$, and $8.59_{-0.11}^{+0.07}$
for IC 4329A, Mrk 335, Mrk 509, Mrk 1239, and PDS 456, respectively (see Table~\ref{tab_mass}).
The other measures of line width yield a similar mass scale once taking into account the associated virial factors (\citealt{Ho2014, Yu2019}).
Here, we adopted $f_{\rm BLR}=1$ with FWHM from the mean spectra for simplicity. Previous calibrations with the $M_\bullet-\sigma_\star$
relation (e.g., \citealt{Ho2014, Yu2019}) reported the virial factor in a range of 0.9-1.2, with a typical scatter of 0.4 dex inherited from the $M_\bullet-\sigma_\star$ relation. We stress that the above black hole estimations do not include the error and intrinsic scatter of the virial factor.
Also, there has been an argument that, as demonstrated by \cite{Dalla2020}, different measures of line width, especially the FWHM from the mean spectrum, might introduce some systematic
biases to the obtained black hole mass. In this sense, caution should be taken when choosing different types of virial products in Table~\ref{tab_mass}.

As mentioned above, \cite{Bentz2023} monitored IC~4329A in 2022, partially overlapping with our season 2022. Their measured H$\beta$ time lag and
line widths are generally consistent with our results within uncertainties, therefore, the virial products are in agreement with our measurements.
For Mrk 335, \cite{Du2019} compiled the measurements of four RM campaigns completed up to 2019 and reported an averaged virial product $\log M_{\rm vir}/M_\odot=6.93_{-0.11}^{+0.10}$ using FWHM from mean spectra. This is also consistent with our results.

There are previous measurements to the host galaxy properties for several of our five targets, from which the black hole mass
can be estimated. Using the $K$-band bulge magnitude estimate, \cite{Vasudevan2010} reported $\log M_\bullet/M_\odot=8.29$ for
IC 4329A and $8.56$ for Mrk 509, with a typical uncertainty of $\sim$0.1 dex. \cite{Grier2013a} measured
the stellar velocity dispersion of Mrk 509 $\sigma_\star=184\pm12~{\rm km~s^{-1}}$, which yields a black hole mass of
$\log M_\bullet/M_\odot=8.33_{-0.16}^{+0.17}$ using the $M_\bullet-\sigma_\star$ relation from \cite{Kormendy2013}.
\cite{Ho2014} estimated a bulge stellar mass of Mrk~335 $\log M_\star/M_\odot\approx9.9$, yielding a black hole mass of
$\log M_\bullet/M_\odot\approx7.41$ using the $M_\bullet-M_\star$ relation from \cite{Kormendy2013}.
If taking into account the systematic scatter of these empirical relations, our results are generally compatible with the above rough estimates.

\section{Discussions}

\begin{figure}[!t]
\centering
\includegraphics[width=0.45\textwidth]{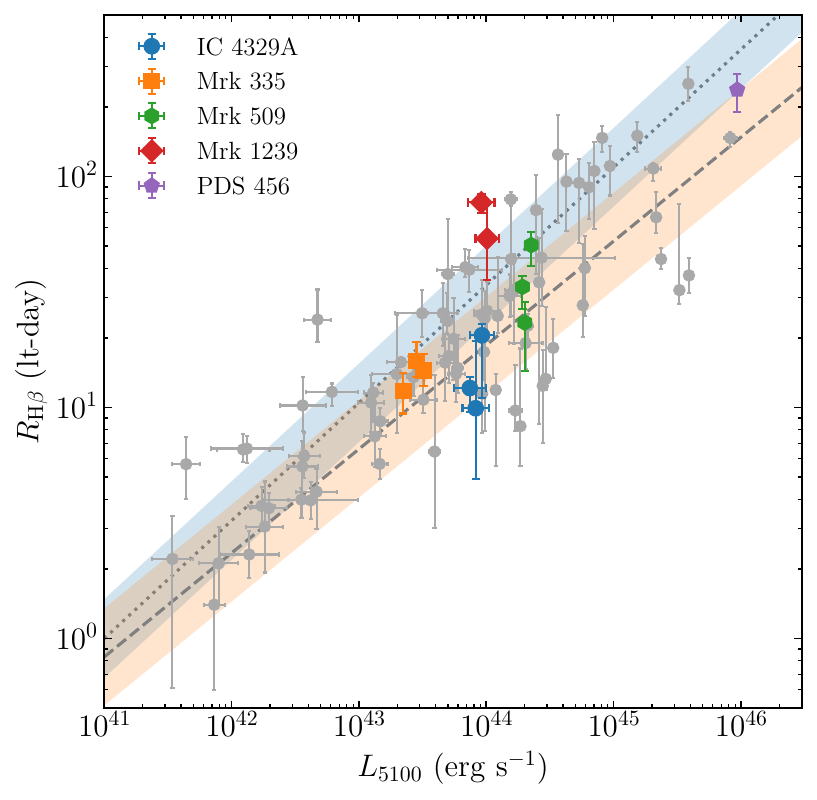}
\caption{The relationship between BLR sizes and AGN 5100~{\AA} luminosities. Grey points are from the compilation by \cite{Du2019} plus the RM campaign by
\cite{Liss2021}. Dotted and dashed lines with shaded bands represent the best fits and scatter for sub- and super-Eddington AGNs from \cite{Du2018}
 (see the text in Section~\ref{sec_rl}).}
\label{fig_rl}
\end{figure}

\subsection{Possible Caveats with Photometric Light Curves}
Instead of using the 5100~{\AA} light curve, we employ the combined photometric light curves in RM analysis because they have much higher cadences. This raises two issues: one is that there might be extra time lags between
the photometric and 5100~{\AA} variations; the other is that the presence of broad emission lines in the filter's
passband will contaminate the continuum variations and thereby affect the time lag measurements.

Regarding the first issue, previous continuum RM studies had illustrated that rest-frame inter-band
continuum time lags follow a power-law relation $\Delta\tau\propto \tau_0[(\lambda/\lambda_0)^{4/3}-1]$, as expected from the standard accretion
disk model (e.g., \citealt{Edelson2019, Guo2022} and references therein), where $\tau_0$ is the time lag (or radius) for photons at the
reference wavelength of $\lambda_0$. The typical value of $\tau_0$ is 0.7-3.6 days for a $10^7$-$10^8M_\odot$ black hole accreting at an Eddington
rate (see Equation 1 in \citealt{Edelson2019}). This results in a time lag of 0.06-0.28 days  between $g$/ZTF-g band and 5100~{\AA}.
Even considering that the observed accretion disk sizes are systematically larger by a factor of $\sim$2 than the anticipated values from
the standard accretion disk model (e.g., \citealt{Edelson2019}), the inter-band time lag is still shorter than one day. Therefore,
the influences to the H$\beta$ time lags can be safely neglected.

For the second issue, our spectral decomposition allows us to estimate the flux contributions of broad emission lines
in the filter's passband. We integrate fluxes of all broad emission line components in the passband of $g$ and ZTF-$g$ filters and
find that broad emission lines contribute roughly $\xi=$13\%-18\% of the total fluxes. The influences to the ICCF can be assessed as follows.
Given two light curves $X$ and $Y$, and $X$ consists of continuum and broad line emissions, namely, $f_X=f_X^c + f_X^l$,
the CCF between $X$ and $Y$ is written
\begin{eqnarray}
&&{\rm CCF}(f_X, f^l_Y) = \frac{E[(f_X-\bar f_X)(f_Y-\bar f^l_Y)]}{\sigma(f_X)\sigma(f^l_Y)}\nonumber\\
&&\qquad=\frac{\sigma(f_X^c)}{\sigma(f_X)}{\rm CCF}(f_X^c, f_Y)+ \frac{\sigma(f_X^l)}{\sigma(f_X)}{\rm CCF}(f_X^l, f^l_Y),
\label{eqn_ccf}
\end{eqnarray}
where $E[x]$ and $\sigma(x)$ represent the expectation and standard deviation of the light curve, respectively, and $\bar f$ represents
the mean flux. If we neglect the time lag between H$\beta$ and other emission lines, Equation~(\ref{eqn_ccf}) demonstrates that
the presence of emission lines in photometric light curves leads to two effects: 1) adding a scaling factor $\sigma(f_X^c)/\sigma(f_X)$
to the CCF without emission line contaminations; 2) adding an additional CCF term that peaks around the zero time lag.
If we further make an approximation that $\sigma(f_X^c)\approx (1-\xi)\sigma(f_X)$ and $\sigma(f_X^l)\approx \xi\sigma(f_X)$,
Equation~(\ref{eqn_ccf}) can be simplified into
\begin{eqnarray}
{\rm CCF}(f_X, f^l_Y) = (1-\xi){\rm CCF}(f_X^c, f_Y) + \xi {\rm CCF}(f_X^l, f^l_Y).
\end{eqnarray}
We expect that the CCF is still dominated by the first term in the right-hand side of Equation~(\ref{eqn_ccf}) and the
influences to H$\beta$ time lags are minor.

To test the above simplified assessment, we directly measure the time lags between the H$\beta$ and 5100~{\AA} light curves of Mrk 335 and obtain H$\beta$ time lags of
$9.07_{-1.51}^{+1.77}$, $13.12_{-4.25}^{+3.42}$, and $13.83_{-2.08}^{+2.68}$ days for the three seasons.
As can be seen, these values are consistent within uncertainties with the measurements listed in Table~\ref{tab_lag}, indicating that the above
assessment is reasonable.

\subsection{The Relationship between BLR Sizes and AGN Luminosities}\label{sec_rl}
Figure~\ref{fig_rl} shows the distributions of the five objects on the relationship between BLR sizes and AGN 5100~{\AA} luminosities ($R_{\rm H\beta}-L_{5100}$)
and makes a comparison with the previous AGN RM sample compiled by \cite{Du2019} plus two luminous AGNs observed by \cite{Liss2021}.
In the H$\beta$ RM sample observed so far,
PDS 456 turns out to be one of the most luminous AGN, with a 5100~{\AA} luminosity of $\log \left(L_{\rm 5100}/{\rm erg~s^{-1}}\right)=45.97\pm0.03$,
comparable to $45.92\pm0.05$ of 3C 273.

Figure~\ref{fig_rl} also superimposes the best fits of the $R_{\rm H\beta}-L_{5100}$ relation for sub- and super-Eddington AGNs from \cite{Du2018}, which were
classified by the dimensionless accretion rate $\mathscr{\dot M}<3$ and $\mathscr{\dot M}\geqslant3$, respectively. Here, the dimensionless accretion rate
is defined as (e.g., \citealt{Du2014})
\begin{equation}
\mathscr{\dot M} = 20.1 \left(\frac{\ell_{44}}{\cos i} \right)^{3/2} m_7^{-2},
\end{equation}
where $\ell_{44}=L_{5100}/10^{44}~\rm erg~s^{-1}$, $m_7=M_\bullet/10^{7}M_\odot$, and $i$ is the inclination angle of the AGN.
Using the estimated black hole mass from the FWHM of mean spectra (with $f_{\rm BLR}=1$), the dimensionless accretion rate is
$\mathscr{\dot M}\approx0.2$, $6.4$, $1.0$, $3.9$, and $18.8$ for IC4329A, Mrk 335, Mrk 509, Mrk 1239, and PDS~456, respectively.
Here, we simply adopt $\cos i=0.75$, an averaged value for the inclination randomly distributed between 0-60$^\circ$.
Overall, the five objects follow the $R_{\rm H\beta}-L_{5100}$ relationship of the previous RM sample
and do not exhibit significant deviations. The two super-Eddington objects Mrk 335 and PDS 456 align marginally with both sub- and super-Eddingtion
relationships. This is because, as demonstrated by \cite{Du2019}, the deviation between two relationships becomes significant when the accretion rate
reaches about $\mathscr{\dot M}\sim100$. Mrk 1239 seems to be an outlier. However, Mrk 1239 has a heavy extinction (\citealt{Marziani1992, Mehdipour2018, Rodriguez-Ardila2006, Pan2021}) and therefore its luminosity is still  subject to some uncertainties.

\subsection{The Virial Factors}
\begin{figure}[t!]
\centering
\includegraphics[width=0.45\textwidth]{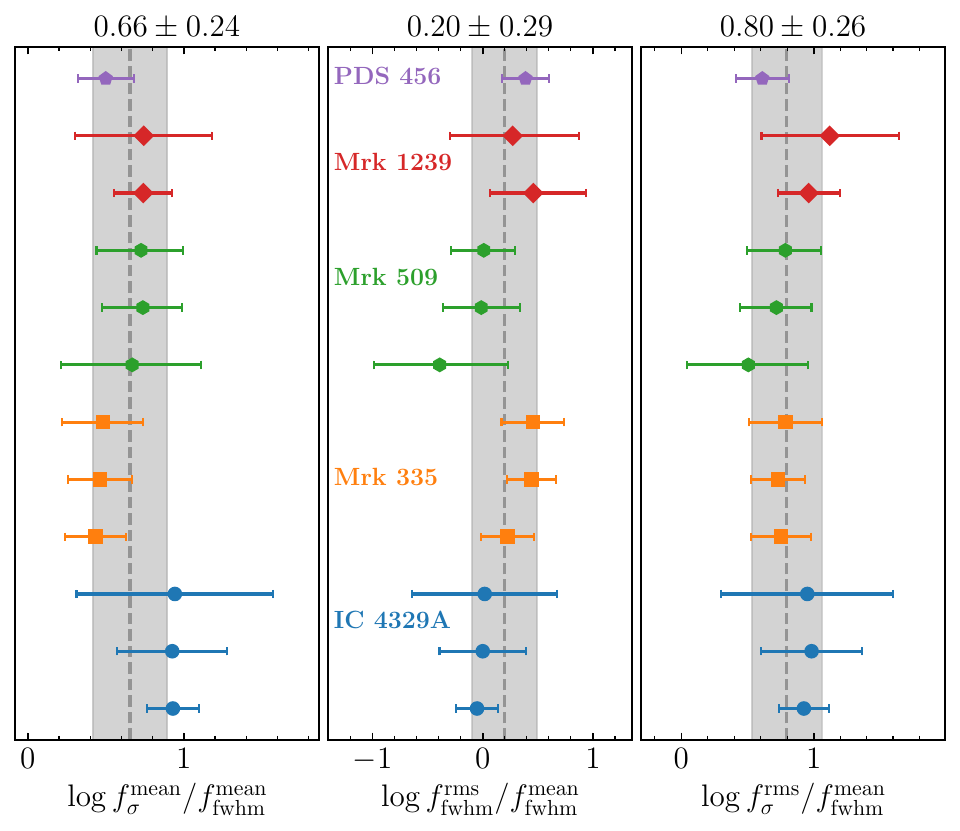}
\caption{The ratios of virial factors for different measures of the H$\beta$ line width with respect to that for H$\beta$ FWHM of the mean spectrum.
From left to right are line dispersion of the mean spectrum, FWHM and line dispersion of the rms spectrum, respectively.}
\label{fig_ratio}
\end{figure}

As mentioned above, to convert the virial product to true black hole mass, a virial factor $f_{\rm BLR}$
is invoked (see Equation~\ref{eqn_mass}), which depends on the adopted measure of line width. While we cannot determine those factors
in this work without independent measurements of the true black hole mass),
we can still derive the ratios between them.
For two types of line widths, e.g., say $V_1$ and $V_2$, the ratio of the corresponding virial factors $f_1$ and $f_2$ is
\begin{equation}
\frac{f_1}{f_2} = \frac{M_{\rm vir, 2}}{M_{\rm vir, 1}} = \frac{V_2^2}{V_1^2}.
\end{equation}
Figure~\ref{fig_ratio} illustrates the ratios for different measures of the H$\beta$ line width with respect to that for H$\beta$ FWHM of the mean spectrum.
We can find that the ratios of different objects are roughly distributed around a common value.
By averaging over the whole measurements, we obtain ratios $\log \left(f_{\sigma}^{\rm mean}/f_{\rm fwhm}^{\rm mean}\right)=0.66\pm0.24$, $\log \left(f_{\rm fwhm}^{\rm rms}/f_{\rm fwhm}^{\rm mean}\right)=0.20\pm0.29$, and $\log \left(f_{\sigma}^{\rm rms}/f_{\rm fwhm}^{\rm mean}\right)=0.80\pm0.26$.
These values are consistent with the calibrations of \cite{Ho2014}  using the $M_\bullet-\sigma_\star$ relation over a sample of AGNs with RM measurements
(see also \citealt{Yu2019}).

It is worth mentioning that BLR dynamical modeling can directly determine the black hole mass of individual AGNs and therefore is able to infer the virial factor object by object
(\citealt{Pancoast2011, Pancoast2014, Li2013, Li2018}). Previous studies in this direction demonstrated that the virial factor might not
be a constant and most likely depends on AGN properties, such as black hole mass, accretion rate, etc
(\citealt{Pancoast2014b, Grier2017, Li2018, Williams2018, Villafana2022}).
Considering more thorough constraints on BLR structure and kinematics with SARM
analysis, we expect an improved determination to the virial factors over a sizeable AGN sample and from which the empirical relation between the virial factor and AGN properties can be better established.

\subsection{Host Galaxy Decomposition from the HST Images}
In our spectral decomposition (see Section~\ref{sec_decomp}), we include a starlight component to eliminate the host galaxy contamination
to the observed 5100~{\AA} flux density. As an independent check, we can also estimate the host galaxy flux density by decomposing the
HST images with \textsc{Galfit} (\citealt{Peng2010}).

We retrieved from the HST archives and found available image data for four of our targets (except for PDS 456). The decomposition procedure
is quite standard. We reduce the science images following the HST instrumental handbook and remove cosmic rays in individual science images with the {\textsc{L.A.Cosmic}} package (\citealt{Dokkum2001}). All frames are then distortion-corrected
and combined with the {\textsc{AstroDrizzle}} task (version 3.3.0). The sky subtraction is not included to preserve the host morphology in faint frames.
From each exposed frame, a point-spread function (PSF) is created at the same detector position with the AGN using the {\textsc{TinyTim}} package (\citealt{Krist2011}). These PSFs are combined with the {\textsc{AstroDrizzle}} configuration the same way as used for the science image.

We fit the 2D surface brightness profile with the PSF to mimic the AGN,
one S\'{e}rsic profile to model the bulge, another S\'{e}rsic profile to model the galactic disk, and a constant value to model the sky background. The S\'{e}rsic profile is expressed as
\begin{equation}
\Sigma(r)=\Sigma_{\rm e}\exp\left\{-\kappa\left[\left(\frac{r}{r_{\rm e}}\right)^{1/n}-1\right]\right\},
\end{equation}
where $\Sigma_{\rm e}$ is the surface brightness at the effective radius $r_e$, $n$ is the S\'{e}rsic index, and $\rm \kappa$ is coupled with $n$ to make sure the effective radius equals to the half light radius.
For Mrk 509, an extra ring component is added to fit the ring-like structure in the residual.
For Mrk~1239, the fluxes in the nuclear region ($\lesssim$0.1\arcsec) are saturated and we therefore mask the inner {0.8\arcsec} nuclear region.
In Appendix~\ref{sec_hst}, we show the decomposition results and the 1D surface brightness profiles.

After fitting the surface brightness profile, we subtract the AGN and sky components to estimate the host flux contamination. Considering the high spatial resolution of the HST images, we simply
blur the AGN-sky-free images with the typical seeing of our spectral observations. We then integrate the fluxes of the blurred images within our spectral extraction
aperture and proportionally remove the background flux, which is estimated by integrating the fluxes with the background aperture. A color correction arising from the difference
between the effective wavelength of the HST filter and rest-frame 5100~{\AA} is included using the bulge-template from \cite{Kinney1996}.
We note that this spectrum is not the same as the host galaxy templates adopted in the spectral decomposition.
However, we find that the difference in spectral shapes is minor and therefore the influence to the color correction is negligible.


The contributions of the host galaxies to the observed 5100~{\AA} flux densities (in units of $10^{-15}~\rm erg~s^{-1}~cm^{-2}~\text{\AA}^{-1}$) we obtained are $23.90\pm0.39$, $1.18\pm0.14$, $1.25\pm0.15$, and $7.11\pm0.83$ for IC 4329A, Mrk 335, Mrk 509, and Mrk 1239, respectively.
As a comparison, our spectral decomposition yields $19.90\pm3.13$, $0.54\pm0.19$, $9.36\pm4.09$
for IC 4329A, Mrk 335, and Mrk 1239. For the other two  objects, Mrk 509 and PDS 456, the spectral decomposition cannot reliably model
the host galaxy component because of its weakness there compared to AGN component. For Mrk 509, using the estimate from the HST image,
the host galaxy contribution is less than 10\% in the observed 5100~{\AA} flux. These comparisons illustrate
that the estimates of host galaxy contaminations from two approaches are generally consistent and also reinforce the validity of
our spectral decomposition.

\subsection{Implication for SA Observations}
Using the obtained H$\beta$ BLR size, we can estimate the BLR angular size of near infrared broad emission lines by assuming that they share the same
BLR size as the H$\beta$. For IC 4329A, Mrk 335, Mrk 509, and Mrk 1239, the broad emission line observable by GRAVITY is Br$\gamma$ (2.166~$\mu$m), while for PDS 456, the observable line is Pa$\alpha$ (1.875~$\mu$m).
In practice, the observed spectral flux always consists of broad line emissions plus the underlying continuum.
As a result, the {\it maximum} angular photocenter offset of the BLR  can be estimated as
\begin{equation}\label{eqn_ang}
\Delta \theta = \frac{c\tau_{\rm BLR}}{D_{\rm A}},
\end{equation}
where $D_A$ is the angular diameter distance. In practice, due to the contribution of the underlying continuum emissions,
the observed photocenter offset is indeed (e.g., \citealt{Li2023})
\begin{equation}\label{eqn_ang_obs}
\Delta \theta_{\rm obs} = \frac{f_{\rm line}}{1+f_{\rm line}} \frac{c\tau_{\rm BLR}}{D_{\rm A}},
\end{equation}
where $f_{\rm line}$ is the flux ratio of the line profile peak to the underlying continuum.
We adopt the typical value of $f_{\rm line}=0.06$ for Br$\gamma$ line and $0.6$ for Pa$\alpha$ line.
The calculated $\Delta \theta$ and $\Delta \theta_{\rm obs}$ are listed in Table~\ref{tab_lag}.
The maximum angular photocenter offset
of the Br$\gamma$ line is $\Delta \theta_{\rm obs}\approx$2.9, 1.4, 3.4, 8.7 $\mu$as for IC 4329A, Mrk 335, Mrk 509, and Mrk 1239, respectively.
For PDS 456, the maximum angular photocenter offset of the Pa$\alpha$ line is $\Delta \theta_{\rm obs}\approx$23 $\mu$as.
These values are somehow different from the estimates mentioned in Section~\ref{sec_targets} because we use our observed BLR size and
also include the line strength ratio $f_{\rm line}$.
From these rough estimates, PDS~456 has the largest BLR angular size and is therefore the most appropriate target for SA observations.
However, we stress that there might be differences in BLR sizes of different emission lines, therefore the above estimates can only be regarded as
an approximation. The more precise estimates are beyond the scope of this paper, but might be obtained by comparing the profiles of Br$\gamma$/Pa$\alpha$
and H$\beta$ lines. In addition, Equation~(\ref{eqn_ang}) does not include the geometry and position angle of the BLR, therefore
realistic photocenter offsets might be even smaller than the values estimated above.

Recently, \cite{Gravity2024} reported SA observations of IC~4329A, Mrk 509, and Mrk~1239 in the year 2021 and PDS 456 during the period of 2018-2021.
Their SA modeling yielded a BLR
size of $13.49_{-5.55}^{+3.49}$, $194.98_{-87.83}^{+4.54}$, $58.88_{-48.17}^{+4.21}$, and $309.03_{-180.20}^{+62.51}$ light-days, respectively.
Except for Mrk 509, these results are well consistent within uncertainties with our measured H$\beta$ time lags (note the time lags in Table~\ref{tab_lag} are quoted
in the observed frame). The SA BLR size of Mrk 509 is subject to a large lower error and the difference from our RM measurements is within the 2$\sigma$ uncertainty. However, it is worth mentioning that Mrk 509 shows different long-term variation trends between the continuum and H$\beta$ line
(see Appendix~\ref{app_whole}). A future SARM analysis will help to clarify whether there is a connection between the different trends and the slightly
larger SA BLR size.

\section{Conclusions}
We conduct an optical, multi-year RM campaign targeting bright AGNs with large anticipated BLR angular sizes appropriate for
SA experiments with the GRAVITY/VLTI instrument. This paper reports the first results for five objects, IC 4329A, Mrk 335, Mrk 509,
Mrk 1239, and PDS 456. We obtain more than 130 epochs of spectra with typical cadence of every 3-5 days for IC 4329A, Mrk 335
and Mrk 509 between 2019 and 2023, and about 90 epochs with a typical cadence of every 6 days for Mrk 1239 and PDS 456 between 2018 and 2023.
The spectra have sufficiently good quality allowing us to detect the H$\beta$ time lags with respect the continuum and
perform velocity-resolved time lag analysis of the H$\beta$ profiles.

We employ the spectral decomposition scheme to extract the H$\beta$ profiles from individual spectra and then integrate the
extracted profiles to obtain the H$\beta$ light curves. Such a procedure minimizes the contamination of the other
emission components that might affect the H$\beta$ time lag analysis as well as H$\beta$ line width measurements. The obtained
H$\beta$ time lags of the five objects generally follow the previously established relationship between BLR sizes and
5100~{\AA} luminosities. The velocity-resolved time lags show diverse structures among the five objects, plausibly implying
diverse BLR kinematics. For Mrk 335 with the most previous
RM studies, there is a hint that the BLR kinematics might be evolving in the long term. However, a more thorough, sophisticated
analysis (such as dynamical modeling) is required to reinforce these results.

By combining the H$\beta$ time lags and four types of line widths, we calculate the corresponding virial products, which show
good consistency throughout all monitored seasons. Using a unity virial factor ($f_{\rm BLR}=1$) and the H$\beta$ FWHM from mean spectra,
we estimates the black hole mass averaged over the seasons $\log M_\bullet/M_\odot=8.02_{-0.14}^{+0.09}$, $6.92_{-0.12}^{+0.12}$, $8.01_{-0.25}^{+0.16}$, $7.44_{-0.14}^{+0.13}$, and $8.59_{-0.11}^{+0.07}$ for IC 4329A, Mrk 335, Mrk 509, Mrk 1239, and PDS 456, respectively.
The black hole mass estimations  using other line width measures are also reported (up to the virial factors) and yield
a similar mass scale.
We also estimate the BLR angular size of near infrared broad emission lines (Br$\gamma$ and Pa$\alpha$) by assuming
that these lines share the same BLR as the H$\beta$ line. We find that PDS~456 has the largest BLR angular size (of the Pa$\alpha$ line)
and therefore is an appropriate target for conducting SA experiments.

\section*{acknowledgements}
J.M.W. thanks E. Sturm for sharing the GRAVITY-targeted AGN list, which allowed to start the present SARM campaign, and also thanks him for sharing the manuscript on the GRAVITY observations of broad-line regions prior to publication.
We thank P. Marziani for useful discussions on the spectral decomposition of IC 4329A.
We thank the support of the staff of the Lijiang 2.4 m telescope and of the CAHA 2.2m telescope.
This work is based on observations collected at the Centro Astron\'omico Hispanoen Andaluc\'ia
(CAHA) at Calar Alto, operated jointly by the Andalusian Universities and the Instituto de Astrof\'isica de Andaluc\'ia (CSIC).
This paper uses observations made from the South African Astronomical Observatory (SAAO) and we thank Francois van Wyk, who carried out most of the observations there.
We acknowledge financial support from the National Key R\&D Program of China (2021YFA1600404 and 2023YFA1607904), the National Natural Science
Foundation of China (NSFC; 11833008, 11991050, and 12333003).
Y.R.L. acknowledges financial support from the NSFC through grant No. 12273041 and from the Youth Innovation Promotion Association CAS.
C.H. acknowledges financial support from NSFC grants NSFC-12122305.
P.D. acknowledges ﬁnancial support from NSFC grants NSFC-12022301 and 11991051.
L.C.H. acknowledges financial support from the NSFC (11721303, 11991052, 12011540375, and 12233001), the National Key
R\&D Program of China (2022YFF0503401), and the China Manned Space Project (CMS-CSST-2021-A04, CMS-CSST-2021-A06)

This work used archival data obtained with the Samuel Oschin Telescope 48-inch and the 60-inch Telescope at the Palomar
Observatory as part of the Zwicky Transient Facility project. ZTF is supported by the National Science Foundation under Grants
No. AST-1440341 and AST-2034437 and a collaboration including current partners Caltech, IPAC, the Weizmann Institute for
Science, the Oskar Klein Center at Stockholm University, the University of Maryland, Deutsches Elektronen-Synchrotron and
Humboldt University, the TANGO Consortium of Taiwan, the University of Wisconsin at Milwaukee, Trinity College Dublin,
Lawrence Livermore National Laboratories, IN2P3, University of Warwick, Ruhr University Bochum, Northwestern University and
former partners the University of Washington, Los Alamos National Laboratories, and Lawrence Berkeley National Laboratories.
Operations are conducted by COO, IPAC, and UW.

The HST image data used in this paper were obtained from the Mikulski Archive for Space
Telescopes (MAST) at the Space Telescope Science Institute. The specific observations analyzed can
be accessed via \dataset[10.17909/en4e-c480]{https://doi.org/10.17909/en4e-c480}.

\software{{\textsc{PyCALI}} (\citealt{Li2024pycali}), {\textsc{MICA}} (\citealt{Li2024mica}), {\textsc{CDNest}} (\citealt{Li2023cdnest})}

\appendix
\section{Time Lag Analysis with All Seasons} \label{app_whole}
In Section~\ref{sec_lag}, we measure the H$\beta$ time lags for the four objects (except PDS~456) in each season separately, but not over all seasons as a whole.
The BLR kinematics of those four objects might change with the dynamical timescale on the order of years (see Equation~\ref{eqn_tdyn}), which will affect time lag analysis
over a long time period and produce fake responses. We keep in mind this point when comparing results from all seasons and individual seasons.

\begin{figure*}
\centering
\includegraphics[width=0.95\textwidth]{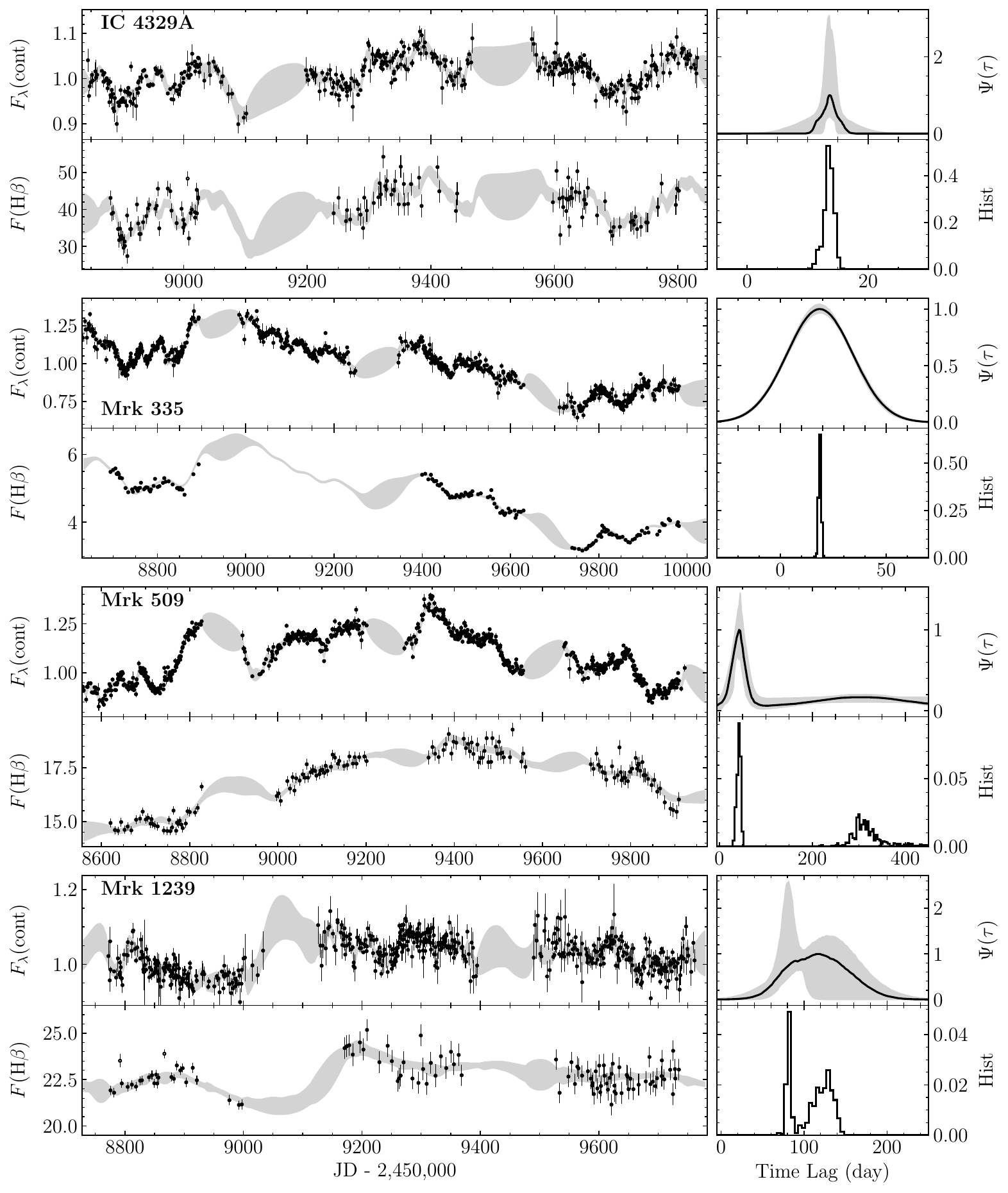}
\caption{The time lag analysis on the full-span light curves using MICA for IC~4329A, Mrk 335, Mrk 509, and Mrk 1239 from top to bottom.
For each object, the left panels show the photometric (top) and H$\beta$ (bottom) light curves. The photometric light curves are converted
from magnitudes and normalized with their means. The H$\beta$ fluxes are in a unit of $10^{-13}~\rm erg~s^{-1}~cm^{-2}$.
The gray shaded bands represent reconstruction. The right panels show
the obtained transfer function (top) and posterior distribution of the time lag (bottom). The gray shaded area represent the uncertainties
of the transfer function. For Mrk 509, two Gaussian components in
the transfer function are needed to appropriately fit the light curves and the time lag distributions shows two separate modes accordingly.}
\label{fig_lag_whole}
\end{figure*}

For the sake of completeness, we present MICA results on the full-span light curves for IC~4329A, Mrk 335, Mrk 509, and Mrk 1239
in Figure~\ref{fig_lag_whole}. The obtained H$\beta$ time lag is $13.43_{-0.79}^{+0.87}$ days for IC~4329A and $18.64_{-0.56}^{+0.58}$ days for Mrk 335.
These values fall in the range of time lags obtained in individual seasons.
For Mrk 1239, the posterior distribution of the H$\beta$ time lag shows two separate peaks, located around 80 and 125 days, respectively.
The longer one might arise from seasonal gaps and we thereby deem the shorter one more reliable.
If only using the distribution mode for the short-lag peak, the H$\beta$ time lag is $81.31_{-2.29}^{+2.66}$ days, close to the measured value in 2019 season.
For Mrk 509, we find that only one Gaussian transfer function is not able to fit the light curves. The analysis
with two Gaussians yields a narrow component peaked at $42.52_{-4.52}^{+3.39}$ days and a very broad component
peaked at $314.88_{-19.87}^{+60.30}$ days. It turns out that the time lag of the narrow component approximately matches those obtained in individual seasons.
The explanation for the second component is not clear, probably stemming from seasonal changes in BLR kinematics or a very extended BLR. Longer monitoring will be useful
to reveal its origin.

\begin{figure*}
\centering
\includegraphics[width=0.8\textwidth]{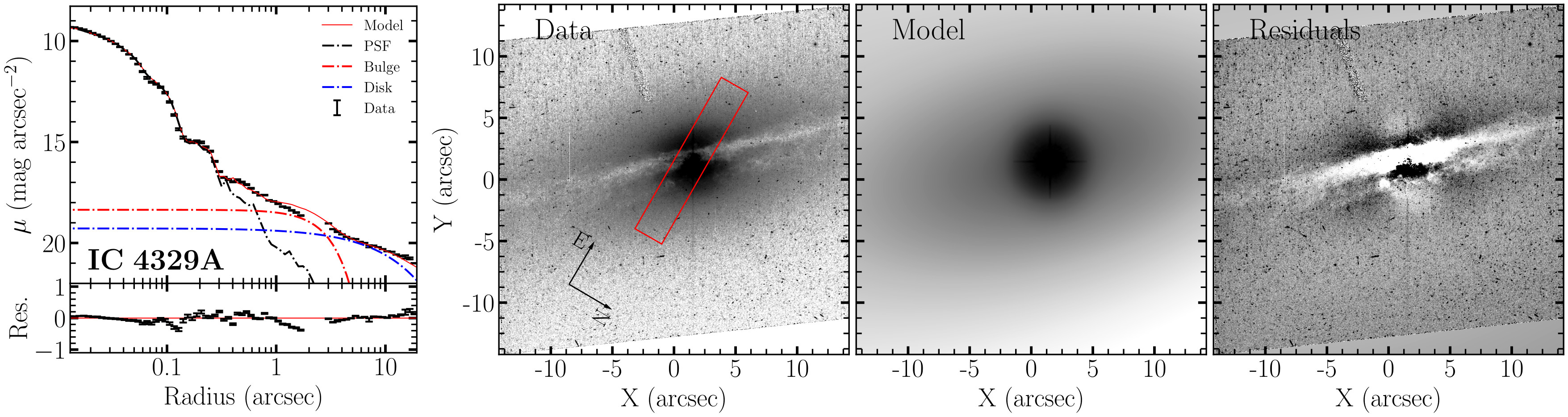}
\includegraphics[width=0.8\textwidth]{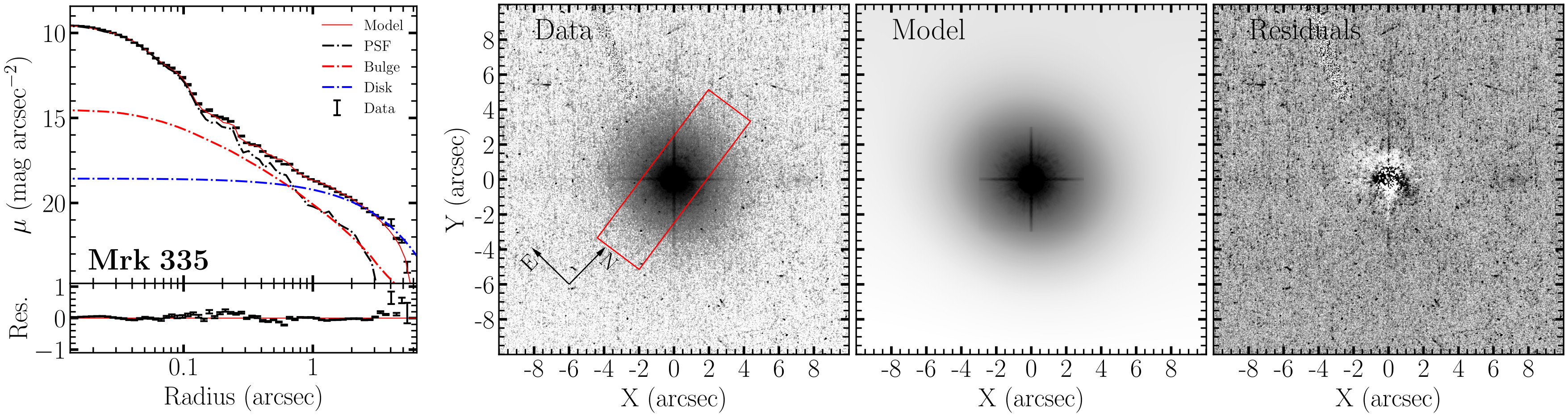}
\includegraphics[width=0.8\textwidth]{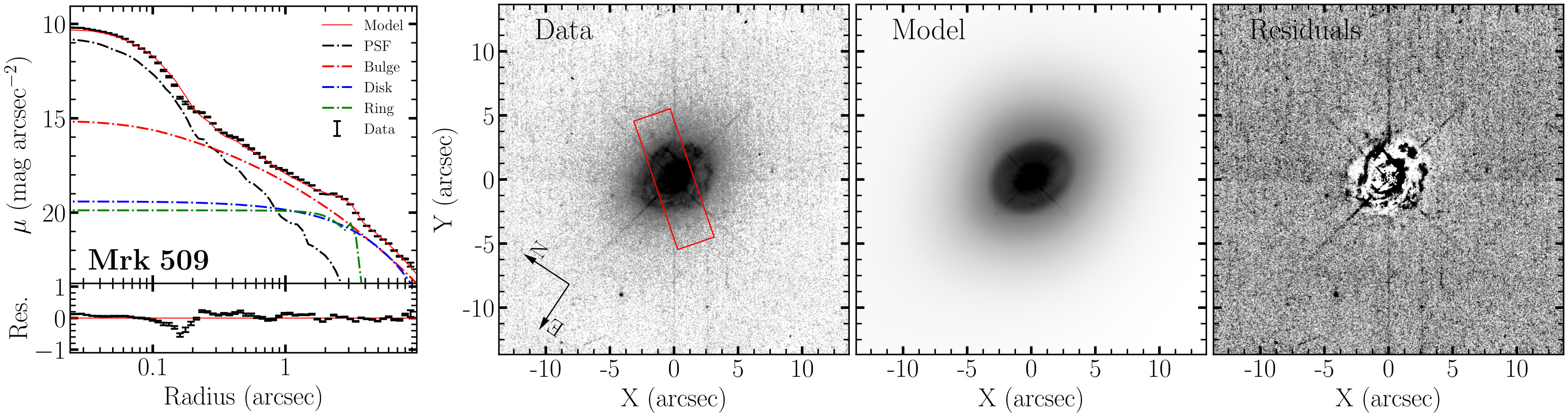}
\includegraphics[width=0.8\textwidth]{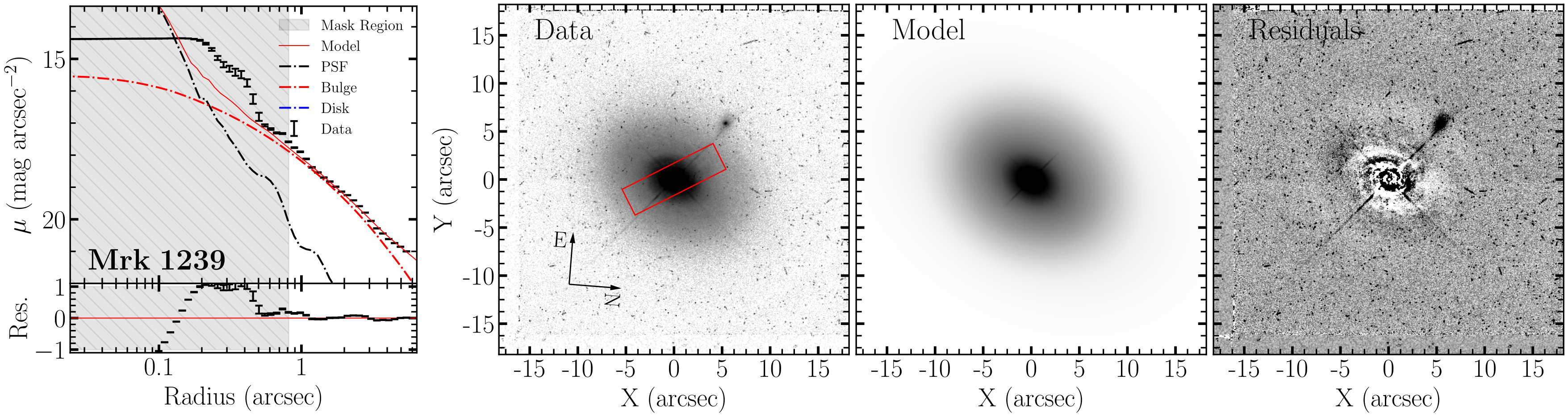}
\caption{HST image decomposition with using \textsc{GALFIT} for IC 4329A, Mrk 335, Mrk 509, and Mrk 1239
from top to bottom. The red rectangle in the second panel from left shows the slit positioning and apertures used for spectral extracting.
The central region ($\leqslant$0.8\arcsec) of Mrk~1239 is masked due to the flux saturation.}
\label{fig_hst}
\end{figure*}

\section{The HST Image Decomposition}\label{sec_hst}
In this appendix, we shows the HST image decomposition results using \textsc{GALFIT} for IC 4329A, Mrk 335, Mrk 509, and Mrk 1239
(see Figure~\ref{fig_hst}). There is no available HST image for PDS 456.

\section{The Light Curve Data} \label{sec_data}
We tabulate the photometric, 5100~{\AA} continuum, and H$\beta$ light curve data in Table~\ref{tab_lc_data}.

\begin{rotatetable*}
\begin{deluxetable}{ccccccccccccc}
\tabletypesize{\footnotesize}
\tablecaption{H$\beta$ Line Widths and Black Hole Mass Measurements\label{tab_mass}}
\tablehead{\colhead{~~~~Name~~~~} & ~~~~Season~~~~ & \multicolumn{4}{c}{Mean} & & \multicolumn{4}{c}{RMS} & ~~~~~$\log L_{5100}$~~~~~ & ~~~~~$\log L_{\rm H\beta}$~~~~~\\\cline{3-6}\cline{8-11}
& & FWHM  & $\log M_{\rm vir}$  & $\sigma$ & $\log M_{\rm vir}$ & & FWHM  & $\log M_{\rm vir}$  & $\sigma$ & $\log M_{\rm vir}$\\
& & (km~s$^{-1}$)  &  ($M_\odot$) & (km~s$^{-1}$) &  ($M_\odot$) & & (km~s$^{-1}$) &  ($M_\odot$) & (km~s$^{-1}$) &  ($M_\odot$) & ($\rm erg~s^{-1}$) & ($\rm erg~s^{-1}$)
}
\startdata
 IC 4329A &  2020 & $6473\pm107$ & $8.00_{-0.11}^{+0.07}$ & $2219\pm16$ & $7.07_{-0.10}^{+0.06}$ & &$6871\pm211$ & $8.05_{-0.12}^{+0.08}$ & $2230\pm61$ & $7.07_{-0.12}^{+0.08}$& $43.87\pm0.12$ & $42.36\pm0.05$\\
         &  2021 & $6538\pm104$ & $8.23_{-0.29}^{+0.07}$ & $2252\pm14$ & $7.31_{-0.28}^{+0.06}$ & &$6536\pm358$ & $8.23_{-0.33}^{+0.10}$ & $2106\pm84$ & $7.25_{-0.31}^{+0.09}$& $43.97\pm0.10$ & $42.44\pm0.05$\\
         &  2022 & $6736\pm94$ & $7.94_{-0.32}^{+0.31}$ & $2275\pm15$ & $7.00_{-0.32}^{+0.31}$ & &$6597\pm240$ & $7.93_{-0.34}^{+0.33}$ & $2252\pm63$ & $6.99_{-0.34}^{+0.32}$& $43.92\pm0.10$ & $42.40\pm0.05$\\
         &   All & \nodata & $8.02_{-0.14}^{+0.09}$ & \nodata & $7.09_{-0.13}^{+0.08}$ & &\nodata & $8.06_{-0.16}^{+0.11}$ & \nodata & $7.09_{-0.16}^{+0.10}$ & $43.93\pm0.11$ & $42.40\pm0.05$\\
  Mrk 335 &  2019 & $1693\pm77$ & $6.91_{-0.11}^{+0.11}$ & $1026\pm18$ & $6.47_{-0.09}^{+0.08}$ & &$1310\pm93$ & $6.69_{-0.14}^{+0.13}$ & $711\pm38$ & $6.16_{-0.12}^{+0.11}$& $43.51\pm0.04$ & $41.93\pm0.02$\\
         &  2021 & $1754\pm87$ & $6.98_{-0.10}^{+0.13}$ & $1031\pm19$ & $6.52_{-0.08}^{+0.10}$ & &$1056\pm42$ & $6.54_{-0.09}^{+0.12}$ & $756\pm14$ & $6.25_{-0.08}^{+0.10}$& $43.45\pm0.04$ & $41.90\pm0.03$\\
         &  2022 & $1817\pm122$ & $6.88_{-0.17}^{+0.13}$ & $1045\pm26$ & $6.40_{-0.13}^{+0.10}$ & &$1075\pm54$ & $6.42_{-0.15}^{+0.12}$ & $733\pm30$ & $6.09_{-0.14}^{+0.11}$& $43.35\pm0.05$ & $41.78\pm0.03$\\
         &   All & \nodata & $6.92_{-0.12}^{+0.12}$ & \nodata & $6.46_{-0.09}^{+0.09}$ & &\nodata & $6.52_{-0.12}^{+0.12}$ & \nodata & $6.17_{-0.11}^{+0.11}$ & $43.44\pm0.04$ & $41.89\pm0.02$\\
  Mrk 509 &  2019 & $3898\pm519$ & $7.84_{-0.36}^{+0.19}$ & $1804\pm45$ & $7.17_{-0.25}^{+0.10}$ & &$6111\pm1253$ & $8.23_{-0.43}^{+0.24}$ & $2177\pm72$ & $7.33_{-0.26}^{+0.11}$& $44.30\pm0.05$ & $42.65\pm0.01$\\
         &  2020 & $4109\pm486$ & \nodata & $1815\pm39$ & \nodata & &$3723\pm201$ & \nodata & $1590\pm47$ & \nodata & $44.37\pm0.01$ & $42.71\pm0.01$\\
         &  2021 & $4139\pm457$ & $8.23_{-0.19}^{+0.15}$ & $1771\pm36$ & $7.49_{-0.11}^{+0.07}$ & &$4203\pm532$ & $8.24_{-0.21}^{+0.16}$ & $1810\pm63$ & $7.51_{-0.12}^{+0.08}$& $44.35\pm0.04$ & $42.74\pm0.01$\\
         &  2022 & $4173\pm552$ & $8.05_{-0.22}^{+0.15}$ & $1809\pm39$ & $7.33_{-0.12}^{+0.06}$ & &$4127\pm144$ & $8.04_{-0.13}^{+0.07}$ & $1688\pm39$ & $7.27_{-0.12}^{+0.06}$& $44.28\pm0.03$ & $42.71\pm0.02$\\
         &   All & \nodata & $8.01_{-0.25}^{+0.16}$ & \nodata & $7.33_{-0.14}^{+0.07}$ & &\nodata & $8.08_{-0.15}^{+0.10}$ & \nodata & $7.33_{-0.14}^{+0.08}$ & $44.35\pm0.02$ & $42.71\pm0.01$\\
 Mrk 1239 &  2019 & $1409\pm108$ & $7.48_{-0.12}^{+0.10}$ & $601\pm24$ & $6.74_{-0.08}^{+0.07}$ & &$830\pm264$ & $7.02_{-0.38}^{+0.28}$ & $466\pm45$ & $6.51_{-0.13}^{+0.12}$& $43.96\pm0.10$ & $42.34\pm0.01$\\
         &  2020 & $1429\pm107$ & $7.33_{-0.25}^{+0.22}$ & $607\pm23$ & $6.59_{-0.21}^{+0.19}$ & &$1045\pm212$ & $7.06_{-0.38}^{+0.32}$ & $393\pm52$ & $6.21_{-0.30}^{+0.27}$& $44.01\pm0.09$ & $42.36\pm0.02$\\
         &  2021 & $1454\pm137$ & \nodata & $617\pm26$ & \nodata & &$2079\pm100$ & \nodata & $609\pm22$ & \nodata & $43.99\pm0.10$ & $42.35\pm0.01$\\
         &   All & \nodata & $7.44_{-0.14}^{+0.13}$ & \nodata & $6.71_{-0.10}^{+0.09}$ & &\nodata & $7.03_{-0.38}^{+0.30}$ & \nodata & $6.41_{-0.17}^{+0.16}$ & $43.99\pm0.10$ & $42.35\pm0.01$\\
  PDS 456 &   All & $2883\pm9$ & $8.59_{-0.11}^{+0.07}$ & $1622\pm1$ & $8.09_{-0.10}^{+0.07}$ & &$1843\pm72$ & $8.20_{-0.14}^{+0.11}$ & $1426\pm37$ & $7.97_{-0.13}^{+0.09}$& $45.97\pm0.03$ & $44.19\pm0.02$\\
\enddata
\end{deluxetable}
\tablecomments{The 5100~{\AA} luminosity ($L_{\rm 5100}$) has been corrected for the contributions from the host galaxy and \ion{Fe}{2} emissions, which
are estimated from the spectral decomposition. A virial factor is invoked to convert virial products to true black hole masses (see Equation~\ref{eqn_mass}), depending on the adopted measure of line width (e.g., \citealt{Ho2014}). It is also worth noting the possible systematics biases/weasknesses for different line widths
(e.g., see \citealt{Dalla2020}).}
\end{rotatetable*}

\begin{deluxetable}{cccc}\label{tab_lc_data}
\tablecaption{Light Curves}
\tablehead{\colhead{~~~~~~~Object~~~~~~~}  & \colhead{~~~~~~~Measure~~~~~~~}  & \colhead{~~~~~~~~~JD~~~~~~~~~} & \colhead{~~~~~~~~~~Value~~~~~~~~~~}\\
& & (-2,450,000)
}
\startdata
 IC 4329A &      g  & 8120.851 & $14.056\pm0.021$ \\
 IC 4329A &      g  & 8122.599 & $14.133\pm0.017$ \\
 IC 4329A &  5100A  & 8881.444 & $57.328\pm0.219$ \\
 IC 4329A &  5100A  & 8883.452 & $51.830\pm0.169$ \\
 IC 4329A &  Hbeta  & 8881.444 & $43.110\pm0.274$ \\
 IC 4329A &  Hbeta  & 8883.452 & $38.819\pm0.209$ \\
  Mrk 335 &  ZTF-g  & 8259.113 & $14.812\pm0.049$ \\
  Mrk 335 &  ZTF-g  & 8263.934 & $14.691\pm0.040$ \\
  Mrk 335 &  5100A  & 8693.623 & $4.832\pm0.010$ \\
  Mrk 335 &  5100A  & 8699.629 & $4.820\pm0.011$ \\
  Mrk 335 &  Hbeta  & 8693.623 & $5.487\pm0.016$ \\
  Mrk 335 &  Hbeta  & 8699.629 & $5.548\pm0.013$ \\
\enddata
\tablecomments{This table is available in its entirety online. Units are magnitude, $10^{-15}~\rm erg~s^{-1}~cm^{-2}~\text{\AA}^{-1}$,
and $10^{-13}~\rm erg~s^{-1}~cm^{-2}$ for photometric, 5100~{\AA} continuum,  and H$\beta$ light curves, respectively.}
\end{deluxetable}

\end{document}